\documentclass[10pt]{article}
\usepackage{color,array,amsthm,amsmath,amssymb}
\usepackage{graphicx}
\usepackage{multirow}
\usepackage{multicol}
\usepackage{subcaption}
\usepackage{tabularray}
\usepackage{pgfplots}
\usepackage{float}
\usepackage{hyperref}
\usepackage{calc}
\usepackage{cuted}
\usepackage{physics}
\usepackage{bm}
\usepackage{verbatim}
\usepackage[sort&compress,numbers]{natbib}
\pgfplotsset{compat=1.18}

\DeclareMathOperator{\sign}{sgn}

\theoremstyle{remark}
\newtheorem{theorem}{Theorem}[section]
\newtheorem{lemma}{Lemma}
\newtheorem{rem}{Remark}

\newtheorem{defn}{Definition}
\newtheorem{proposition}{Proposition}

\usepackage[switch]{lineno}
\usepackage{url}
\usepackage{xcolor}
\definecolor{newcolor}{rgb}{.8,.349,.1}

\usepackage{indentfirst,csquotes}

\begin{document}

\title{Robust Fuel-Optimal Landing Guidance for Hazardous Terrain using Multiple Sliding Surfaces} %%%%%%%%%%%%
\author{Sheikh Zeeshan Basar, Satadal Ghosh}
\date{\today}
% \address{Address}
% \email{example@mail.com}
\maketitle

\begin{abstract}
%%%
In any spacecraft landing mission, fuel-efficient precision soft landing while avoiding nearby hazardous terrain is of utmost importance. 
Very few existing literature have attempted addressing both the problems of precision soft landing and terrain avoidance simultaneously.
To this end, an optimal terrain avoidance landing guidance (OTALG) was recently developed, which showed promising performance in avoiding the terrain while consuming near-minimum fuel. However, its performance significantly degrades in the face of external disturbances, indicating lack of robustness.
To mitigate this problem, in this paper, a near fuel-optimal guidance law is developed to avoid terrain and achieve precision soft landing at the desired landing site. 
Expanding the OTALG formulation using sliding mode control with multiple sliding surfaces (MSS), the presented guidance law, named `MSS-OTALG', improves precision soft landing accuracy. 
Further, the sliding parameter is designed to allow the lander to avoid terrain by leaving the trajectory enforced by the sliding mode and eventually returning to it when the terrain avoidance phase is completed.
And finally, the robustness of the MSS-OTALG is established by proving practical fixed-time stability. 
Extensive numerical simulations are also presented to showcase its performance in terms of terrain avoidance, low fuel consumption, and accuracy of precision soft landing under bounded atmospheric perturbations, thrust deviations, and constraints. 
Comparative studies against existing relevant literature validate a balanced trade-off of all these performance measures achieved by the developed MSS-OTALG.

\end{abstract}

\section{Introduction}
The key objective of extra-planetary landing missions is to collect samples, analyse them and relay the data back to Earth. The spacecraft needs to carry an extensive set of scientific module, avoid any damage during the entry, descent and landing (EDL) phase and land as close to the area of interest as possible, even under disturbances and uncertainties, to get the most out of these missions (\cite{Mars_2020}). 
{Besides, when the spacecraft is close to the surface and is about to land, it must be able to reliably avoid undulating terrain such as hillocks or cliff faces(\cite{Acikmese_Ploen_2007}).
Hence, hazard-avoided precision soft landing guidance is essential for safe landing of the spacecraft at a desired location. Thus, in this context, studying fuel optimality, terrain avoidance, and robustness in an integrated manner is also crucial for any EDL mission.
In this paper the terrain avoided planetary landing of a spacecraft is considered, the spacecraft is also equivalently referred to as `lander'.}
Fuel optimality and precision soft-landing accuracy have been significantly researched since the first attempt at reaching a celestial body. 
Methods ranging from feedback guidance, nonlinear control, optimal control, convex optimisation, and learning-based methods have been studied in the literature to reach the desired landing site safely and precisely (\cite{Simplício_Marcos_Joffre_Zamaro_Silva_2018, Chai_Tsourdos_Savvaris_Chai_Xia_Philip_Chen_2021}). 
To address the nonconvexities associated with the landing guidance problems, algorithms such as {Lossless Convexification (LCVX) and Successive Convexification (SCVX)} have been developed (\cite{Acikmese_Ploen_2007,Mao_Szmuk_Açıkmeşe_2016}). 
However, convexification-based methods are open-loop and, therefore, highly susceptible to perturbations and estimation errors. 
Another optimisation-based method was proposed by \cite{Swaminathan_U.P_Ghose_2020} where modified Newton–Raphson scheme in the pseudo-spectral global collocation framework was used to generate fuel-optimal trajectories for pinpoint Mars landing in a real-time manner.
Classical feedback laws for missile guidance, like Proportional Navigation Guidance (PNG), and its adaptations, such as biased PNG (\cite{Byung_Jang_Hyung_1998}), have been explored for powered descent. 
The concept of zero-effort-miss (ZEM) is extensively used in missile guidance, which denotes the miss distance from the desired terminal position if no control effort is applied from the current time forward. 
The idea of ZEM was extended by \cite{Ebrahimi_Bahrami_Roshanian_2008} to include the deviation in terminal velocity, zero-effort-velocity (ZEV). 
Then using results from optimal control, Optimal Guidance Law (OGL) was developed as a function of ZEM and ZEV. Using fractional polynomials \cite{Lu_2020} presented a generic class of powered descent guidance laws, two special cases of which were found to be the conventional Apollo lunar guidance law and the classical ZEM/ZEV guidance law. In a recent literature, Deep Neural Networks have also been used to generate planetary landing trajectories based on optimal control-based formulation (\cite{Sánchez-Sánchez_Izzo_2018}). 
Improving upon this, a theory-supported learning method was proposed by \cite{You_Wan_Dai_Rea_2021} to alleviate the lack of theoretical guarantees (vis-a-vis convergence and local optimality), leading to a reduced learning space dimension.
Generally, optimal solutions from classical trajectory optimization routines (such as SNOPT) are used to generate the offline training data, which limits their practical applicability in case of complicated landing scenarios, especially in the vicinity of a hazardous terrain.

Sliding Mode based augmentation of OGL was proposed by \cite{Furfaro_Gaudet_Wibben_Kidd_Simo_2013} to make the system robust against perturbations.
Due to its effectiveness , multiple sliding surfaces (MSS) have also been used to improve the robustness against external disturbances. MSS has attracted much attention for space applications in the recent past. For example, \cite{Furfaro_Cersosimo_Wibben_2013} used MSS for precision landing in asteroids as well as for autonomous landing on Mars as described by \cite{Gong_Guo_Lyu_Ma_Guo_2022}. In the presence of atmospheric disturbances, an optimal sliding guidance (OSG) presented by \cite{Wibben_Furfaro_2016} was found to perform well with high degree of precision for soft landing even with partial loss of thrust.
The guidance laws proposed by \cite{Furfaro_Cersosimo_Wibben_2013,Gong_Guo_Lyu_Ma_Guo_2022} and \cite{Wibben_Furfaro_2016} have been proved to be finite time stable (FTS) as well.

Note that the guidance laws mentioned above either did not consider to avoid the terrain or used simple glideslopes or glideslope-like constraints to avoid crashing into the terrain. More dedicated studies on terrain avoided landing has also been presented in recent literature. For example, \cite{Bai_Guo_Zheng_2020} used the results of {LCVX} to incorporate terrain avoidance in fuel-optimal powered descent phase. But, the proposed algorithm therein is computationally heavy and has an open-loop structure, thus making it susceptible to disturbances and hence lacking in robustness. 
A more general terrain avoidance guidance law was presented by \cite{Wang_Guo_Ma_Wie_2021} where the classical ZEM/ZEV guidance law was split into two phases by terminating the first phase by a virtual velocity point to prevent subsurface constraint violations.
To avoid the terrain, in 2-dimensional setting, a notion of signed curvature was introduced in the second phase.
While the proposed guidance law was able to avoid terrain and land accurately at the desired landing site with low fuel consumption, the terrain avoidance is done in an heuristic manner, and no guarantees for terrain avoidance were presented for successful terrain avoidance in full 3-dimensional setting. 
Using Barrier Lyapunov functions by \cite{Gong_Guo_Ma_Zhang_Guo_2021} and Prescribed Performance functions by \cite{Gong_Guo_Ma_Zhang_Guo_2022}, guidance laws for terrain-avoided soft landing were proposed. Both the guidance laws were able to manoeuvre to avoid the terrain and soft land at the desired landing point, however, both of them required several difficult-to-estimate time-dependent variables to generate the terrain bounding barriers.
Additionally, these approaches did not consider the aspect of fuel efficiency and achieving satisfactory precision performance within thrust constraints while designing the guidance laws.
Unlike the results by \cite{Gong_Guo_Ma_Zhang_Guo_2021} and \cite{Gong_Guo_Ma_Zhang_Guo_2022}, a much simpler yet effective method for generating barrier functions to cover \textit{a-priori} known terrain was presented by \cite{basar2023fueloptimal}, in which polynomials were used as the barrier to bound the terrain (approximated as multi-stepped shapes). Then, the standard 2-norm performance index for control effort was augmented with a penalty function in terms of distance of the lander to the barriers to develop an Optimal Terrain Avoidance Guidance Law (OTALG). It was shown to be near-fuel-optimal with desired precision in landing while avoiding terrain. However, OTLAG was not guaranteed to possess robustness against external disturbances.

For a successful landing, it is necessary that the lander's guidance law has the following features: terrain avoidance and the ability to land softly and precisely near the desired landing site. Furthermore, it is desirable that the guidance law is also robust against disturbances and has a low fuel consumption. However, most of the existing literature address only some subset of these features. 
For example, \cite{Ebrahimi_Bahrami_Roshanian_2008, Furfaro_Gaudet_Wibben_Kidd_Simo_2013, Furfaro_Cersosimo_Wibben_2013,Gong_Guo_Lyu_Ma_Guo_2022,Wibben_Furfaro_2016} addressed the problem of fuel optimality and precision soft landing, but terrain avoidance was not discussed, while \cite{Gong_Guo_Ma_Zhang_Guo_2021, Gong_Guo_Ma_Zhang_Guo_2022} addressed the problem of terrain avoidance but not of the fuel optimality and precision landing. 
On the other hand, the guidance presented by \cite{Acikmese_Ploen_2007, Mao_Szmuk_Açıkmeşe_2016} and \cite{Bai_Guo_Zheng_2020} are optimal and can avoid the terrain, but is not robust against disturbances. 
In resutls presented by \cite{Wang_Guo_Ma_Wie_2021}, the terrain avoidance is not guaranteed for precision soft landing in full 3-dimensional setting. 
To the best of the authors' knowledge, this paper is the first attempt to satisfy all of the aforementioned features in an integrated manner.
To this end, a novel robust guidance law, named MSS-OTALG, is developed in this paper by expanding upon the optimal guidance formulation (OTALG) by \cite{basar2023fueloptimal} and leveraging Multiple Sliding Surfaces (MSS) as described by \cite{Furfaro_Cersosimo_Wibben_2013}. 
The first sliding surface is established to monitor the {position error} relative to the target landing site, with a virtual controller introduced to ensure convergence of this sliding variable, while the second sliding surface is implemented to ensure that the first sliding variable follows the virtual control.
Global finite time convergence to both the sliding surfaces is proved.
To navigate around rough terrain, the lander might have to deviate from the path dictated by the sliding mode control. Consequently, the sliding parameter, which ensures the overall stability of the second sliding surface, might not be ideal for executing terrain avoidance manoeuvres. 
To address this issue, the sliding parameter is suitably varied based on the system's states and time-to-go such that it allows the system dynamics to deviate from the sliding surface to facilitate terrain avoidance.
Furthermore, with this selection of the sliding parameter, the practical fixed-time stability (PFTS) (\cite{Polyakov2012,Jiang_Hu_Friswell_2016} )of the proposed MSS-OTALG is also established.

The rest of the paper is organised as follows. Section \ref{sec:Problem_Preliminaries} provides background on lander kinematics, presents the OGL of \cite{Ebrahimi_Bahrami_Roshanian_2008} in terms ZEM/ZEV, and reiterates the critical results of OTALG by \cite{basar2023fueloptimal}, on which the main results of this paper rely on. Section \ref{sec:RobustOTALG} develops the guidance law using multiple sliding surfaces and presents the robustness analysis. A discussion on the choice of sliding parameter is presented in Section \ref{sec:ChoiceOfPhi}, and a new sliding parameter is defined. Here, the PFTS of MSS-OTALG law guided landing is proved as well. Finally, Section \ref{sec:Simulations} presents the results from extensive numerical simulations.

\section{BACKGROUND AND PRELIMINARIES}\label{sec:Problem_Preliminaries}
\subsection{Dynamics and Preliminary Results}\label{subsec:dynamics}
\begin{figure}[h]
    \centering
    \includegraphics[width = 0.6\linewidth]{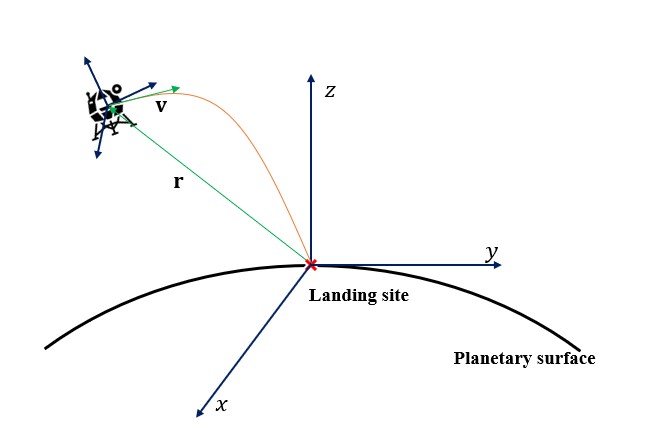}
    \caption{Spacecraft Landing Geometry}
    \label{fig:dynamics}
\end{figure}

{The landing site serves as the origin of a local East-North-Up (ENU) coordinate frame. In this right-handed Cartesian coordinate system, the East axis (X) points eastward parallel to the local latitude line, the North axis (Y) points toward true north parallel to the local longitude line, and the Up axis (Z) completes the orthogonal set by pointing away from the planet's centre normal to the reference ellipsoid surface (\cite{Seeber_2003}) at the landing site, as shown in Fig. \ref{fig:dynamics}. }
% The non-rotating inertial ENU-frame with origin at the landing site is considered as shown in Fig. \ref{fig:dynamics}. 
Assuming a 3-DOF dynamics in $\mathbb{R}^3$ domain, the lander can be modelled as:
\begin{align} \label{eq: dynamics}
\begin{array}{ll}
    &\dot{\mathbf{r}} = \mathbf{v}\\
    &\dot{\mathbf{v}} = \mathbf{a}_c + \mathbf{g} + \mathbf{a}_\mathrm{p}\\
    &{\mathbf{a}} = \frac{\mathbf{T}}{m}\\
    &\dot{m} = -\frac{\Vert \mathbf{T} \Vert}{I_{\mathrm{sp}}g_e}
\end{array}
\end{align}
where $\mathbf{r},\,\mathbf{v}$ represent the position and velocity of lander, and $\mathbf{g}$ is the local gravity, and since the altitude at which the powered descent stage starts is much smaller than the radius of the planet, $\mathbf{g} = [0,0,-g]^\mathrm{T}$ is a valid assumption.
The guidance command is represented by $\mathbf{a}_c$, $\mathbf{a}_\mathrm{p}$ is the net acceleration caused due to bounded perturbations (e.g. wind), $m$ is the lander's mass, $I_{\mathrm{sp}}$ is the specific impulse, and $g_e$ is the gravitational acceleration of Earth. 

The performance index $J = 0.5\int_{t_0}^{t_f} \mathbf{a}^\mathrm{T}\mathbf{a}\,d\mathrm{t}$ was minimised by \cite{Ebrahimi_Bahrami_Roshanian_2008}, subject to \eqref{eq: dynamics} and fixed final time $t_f$ with $\mathbf{a}_\mathrm{p} = 0$, to derive the OGL as:
\begin{align} \label{eq:classical_acc}
    \mathbf{a} = \frac{6}{t_{\mathrm{go}}^2}\mathbf{ZEM} - \frac{2}{t_{\mathrm{go}}}\mathbf{ZEV}.
\end{align}
where $t_{\mathrm{go}} \triangleq t_f - t$ is the time-to-go and:
\begin{align} \label{eq:zeroeffort}
    \begin{array}{ll}
         &\mathbf{ZEM} \triangleq {\mathbf{r}_f^d} - \left[\mathbf{r}(t) + \mathbf{v}t_{\mathrm{go}} + 0.5\mathbf{g}t_{\mathrm{go}}^2\right]  \\
         &\mathbf{ZEV} \triangleq {\mathbf{v}_f^d} - \left[ \mathbf{v}(t) + \mathbf{g}t_{\mathrm{go}} \right]
    \end{array}
\end{align}
here ${\mathbf{r}_f^d},\, {\mathbf{v}_f^d}$ represents desired position and velocity at $t_f$.

\subsection{Optimal Terrain Avoidance Guidance Law}\label{subsec:OTALG}
\begin{figure}
    \centering
    \includegraphics[width = 0.55\linewidth]{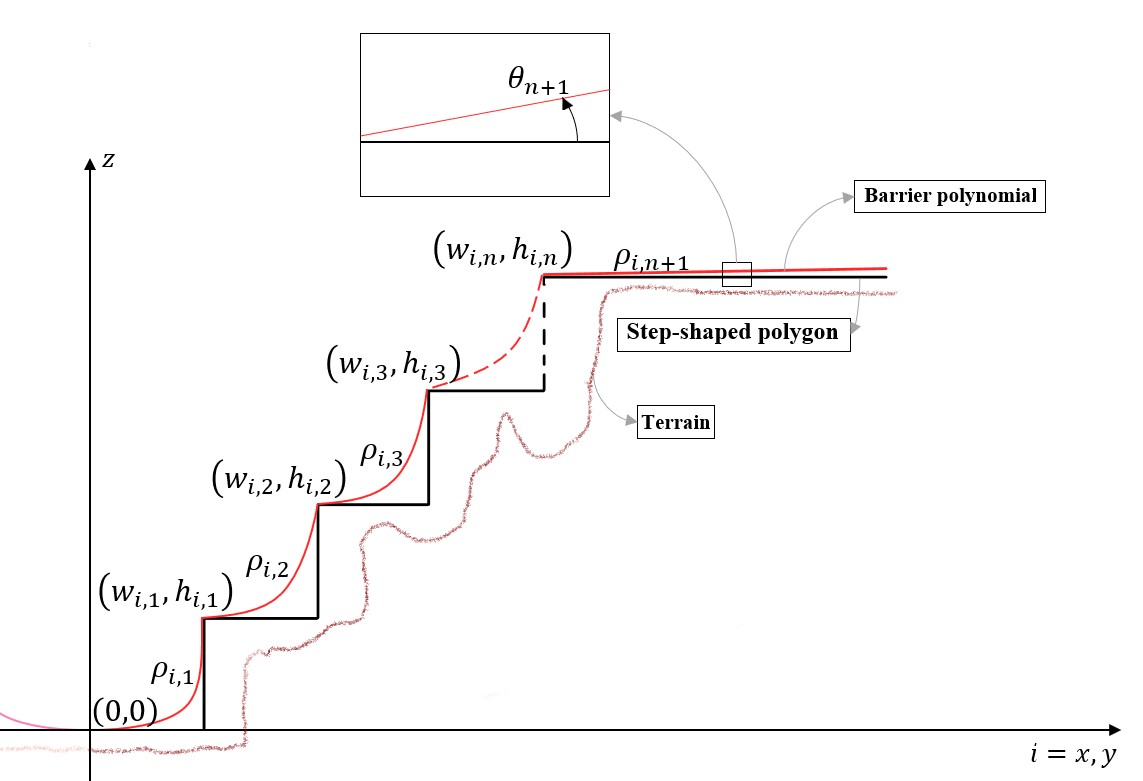}
    \caption{Illustration of terrain and barrier around $n$-step shaped polygons. (\cite{basar2023fueloptimal})}
    \label{fig:barrierPlot}
\end{figure}

To avoid crashing in to the surface, the results by \cite{Ebrahimi_Bahrami_Roshanian_2008} were extended in \cite{Zhou_Xia_2014} and \cite{Zhang_Guo_Ma_Zeng_2017} by introducing a penalty term, which was a function of the distance of lander from the surface, to the standard performance index for fuel optimality. On the other hand, the idea of barriers were introduced by \cite{Gong_Guo_Ma_Zhang_Guo_2021} to avoid more general terrain, however fuel optimality was not considered. A novel penalty to the performance index was introduced by \cite{basar2023fueloptimal}, which was a function of the distance of lander with respect to the barriers, defined as $d_i \triangleq r_i - \rho_{i,j}$ where $i = x,\, y,\, z$. 
{Physically, $d_i$ represents how far the lander is with respect to the barrier surface, and $\rho_{i,j}$ represents the $j$\textsuperscript{th} barrier polynomial \eqref{eq:rho_ij} along $i$\textsuperscript{th} axis. Here, prior obtained terrain information can be used to approximate the terrain as $n$-step shapes and pre-define $n+1$ number of barrier polynomials $\rho_{i,j}$ using the $n$-steps where $j$ is the counter for these $n$ steps. An illustration of the barriers along with the terrain approximated as $n$-step shapes is shown in Fig. \ref{fig:barrierPlot}. The barriers are created as follows, details of which could be found in \cite{basar2023fueloptimal}. For restricting horizontal motion, the polynomials are used to generate the barrier. The first $n$ barriers are polynomials of degree greater than 1, while the $n+1$\textsuperscript{th} barrier is a linear polynomial. The horizontal motion barriers are then defined as:
\begin{equation}\label{eq:rho_ij}
    \rho_{i,j} = \left\{\begin{array}{ll}
         &\pm \left( \beta_{i,j}(r_z + \gamma_{i,j})^{\frac{1}{\lambda_{i,j}}} + \alpha_{i,j}\right),\,h_{i,(j-1)} \leq r_z \leq h_{i,j}\\
         &\pm \left( \beta_{i,(n+1)}(r_z + \gamma_{i,(n+1)}) + \alpha_{i,(n+1)}\right),\,r_z \geq h_{i,n}
    \end{array}\right.
\end{equation}
where $i = x,\,y$ and $j = 1,\, \dots,\, n$.
From the first barrier to the $n^\textsuperscript{th}$ barrier, the constants are denoted as: $\alpha_{i,j} = w_{i,(j-1)};\:\beta_{i,j}= \frac{w_{i,j} - w_{i,(j-1)}}{\left(h_{i,j} - h_{i,(j-1)}\right)^{\frac{1}{\lambda_{i,j}}}};\:\gamma_{i,j} = -h_{i,(j-1)}$
and $\lambda_{i,j}$ is a positive, even natural number. Note that the height of $j^\textsuperscript{th}$ step is defined as $h_{i,j}$, and the horizontal distance from the landing site (origin) along $i$-axis as $w_{i,j}$, with $h_{i,0} = 0$, and $w_{i,0} = 0$. For the $(n+1)^\textsuperscript{th}$ barrier, we first choose the slope angle of the barrier, $\theta_{(n+1)}$ with respect to the axis under consideration. The angle can be chosen as a small value (approx. $0.05^{\circ}\,-\,0.1^{\circ}$) for a relatively flat landing site, or a higher value (approx. $5^{\circ}\,-\,10^{\circ}$) if there is hillock near the landing site. The constants can now be defined as $\alpha_{i,(n+1)} = w_{i,n};\:\gamma_{i,(n+1)} = -h_{i,n}\:\beta_{i,(n+1)} = \tan{\left({\pi}/{2} - \theta_{(n+1)} \right)}$.}

{Now, to restrict lander motion in vertical direction, a small margin, $\delta$, is added to the height of the next lower step to generate the barriers. However, here we run into a problem that if the lander is within the lateral bound of $j^\textsuperscript{th}$ step, but is above the height of $(j+1)^\textsuperscript{th}$ step, the lander would keep on bouncing off the vertical barrier corresponding to the $(j+1)^\textsuperscript{th}$ step. To address this issue, we select the vertical barrier using the following simple comparison: 
\begin{equation}
    \rho_{z} = \left\{\begin{array}{ll}
    h_{i,n} + \delta, & r_z \geq h_{i,n}\\
    h_{i,(j-1)} + \delta, & \left( h_{i,(j-1)} \leq r_z \leq h_{i,j} \right)\ \mathrm{AND}\\
                  &\left( w_{i,(j-1)} \leq \Vert [r_x,\, r_y]^\mathrm{T} \Vert_{\infty} \leq w_{i,j} \right)\\
    % h_{i,(j-2)} + \delta, & \left( h_{i,(j-1)} \leq r_z \leq h_{i,j} \right)\ \mathrm{AND}\\
    %               &\left( w_{i,(j-1)} \leq \Vert [r_x,\, r_y]^\mathrm{T} \Vert_{\infty} \right)
    \end{array}\right.
\end{equation}
where, $j = 1,\,\dots,\,n$.}

% The figure also details the expressions used to create the barrier polynomials. 
% The constants $\alpha_{i,j}$, $\beta_{i,j}$ and $\gamma_{i,j}$ are determined using the height and width of each step w.r.t the origin. Further details of the barrier formulation can be found in Section 3 of the paper by \cite{basar2023fueloptimal}. 
{The modified performance index in this paper by \cite{basar2023fueloptimal} is considered as:
\begin{equation}
    J = 0.5\int_{0}^{t_f} \big[ \mathbf{a}^\mathrm{T}\mathbf{a} - \sum_{i} l_{3,i}e^{-\psi_i} \big]\, d\mathrm{\tau}
\end{equation}
where $e^{-\psi_i}$ is the augmentation term with $\psi_i \triangleq {l_{2,i}}/{(d_i^2 + l_{1,i})}$, and here $l_{1,i},\,l_{2,i},\,l_{3,i} > 0$ are constants. Solving the minimum control effort problem, we get the near-fuel-optimal terrain avoidance guidance law as:
\begin{align}\label{eq:aOptimal}
    \mathbf{a}_{\mathrm{OTALG}} = \frac{6}{t_{\mathrm{go}}^2}\mathbf{ZEM} - \frac{2}{t_{\mathrm{go}}}\mathbf{ZEV} + \mathbf{p}\frac{t_{\mathrm{go}}^2}{12},
\end{align}
where $\mathbf{ZEM}$ and $\mathbf{ZEV}$ are as defined in \eqref{eq:zeroeffort}. Comparing \eqref{eq:classical_acc} and \eqref{eq:aOptimal}, $ \mathbf{p} t_{\mathrm{go}}^2/12$ is responsible for the divert manoeuvre and $\mathbf{p} \triangleq [\dot{p}_{r_x},\, \dot{p}_{r_y},\, \dot{p}_{r_z}]$ where:
\begin{align}
    \dot{p}_{r_i} = l_{2,i}l_{3,i}\frac{d_i\mathrm{e}^{-\psi_i}}{( d_i^2 + l_{1,i} )^2}. \label{eq: p_i}
\end{align}
When the lander is far away from the barriers, the augmentation is large and positive, so the term inside the integration is small. When the lander is close to the barrier, the augmentation is almost zero, increasing the cost, thus generating an acceleration command in direction opposite to the direction of motion to avoid crashing into terrain.}

\section{ROBUST OTALG USING MULTIPLE SLIDING SURFACES (MSS)}\label{sec:RobustOTALG}
The OTALG presented by \cite{basar2023fueloptimal} showed good performance in terms of near-fuel-optimality and terrain avoidance compared to the existing literature. While OTALG is able to avoid the terrain with near-fuel-optimality, it was lacking in terms of accuracy of precision soft landing. Further, it was not designed to reject disturbances. Hence, expanding on the OTALG, multiple sliding surfaces are used here to improve its robustness and guarantee precision soft landing while having low fuel consumption even in the face of bounded external disturbances.

\subsection{MSS Design: Surface 1}\label{subsec:MSS1}
The first sliding surface is defined as $\mathbf{s}_1 \triangleq \mathbf{r} - {\mathbf{r}_f^d}$ where ${\mathbf{r}_f^d}$ is the desired terminal position. This sliding surface is used to track the position error. To track $\mathbf{s}_1$ and drive the position error to zero, a virtual controller is defined as $\Dot{\mathbf{s}}_1 = -\frac{\Lambda}{t_{\mathrm{go}}}\mathbf{s}_1$ where $\Lambda > 0$ is constant.

\begin{theorem} \label{thm: s1_stability}
    \textnormal{The virtual controller defined by $\Dot{\mathbf{s}}_1 = -\frac{\Lambda}{t_{\mathrm{go}}}\mathbf{s}_1$, is globally stable. Further, both $\mathbf{s}_1$ and its derivative, under the reaching law defined by the virtual controller reach zero in finite time.}
\end{theorem}
\begin{proof}
    Global stability can be guaranteed for the virtual controller using Lyapunov's second method, and choosing the candidate Lyapunov function as $V_1 = 0.5\mathbf{s}_1^\mathrm{T}\mathbf{s}_1$. From direct observation, it is clear that $V_1$ is positive definite everywhere, except at $\mathbf{s}_1 = \mathbf{0}$ where $V_1(\mathbf{s}_1 = \mathbf{0}) = 0$. Further, the candidate Lyapunov function is radially unbounded. To guarantee global stability, the time derivative of $V_1$ must be negative definite everywhere.
    {Now considering the virtual controller $\Dot{\mathbf{s}}_1 = -\frac{\Lambda}{t_{\mathrm{go}}}\mathbf{s}_1$, we get:}
    % we get V1_dot == {} which gives V1_dot < 0.  
    % , that is $\Dot{V}_1 = \mathbf{s}_1^\mathrm{T}\Dot{\mathbf{s}}_1 < 0$. This implies:
    
    \begin{align}
        % \Dot{V}_1 &= \mathbf{s}_1^\mathrm{T}\Dot{\mathbf{s}}_1 \nonumber\\
        \Dot{V}_1 &= -\frac{\Lambda}{t_{\mathrm{go}}}\mathbf{s}_1^\mathrm{T}\mathbf{s}_1 < 0 \label{eq: V1dotNeg}
    \end{align}
    {Equation \eqref{eq: V1dotNeg} proves that the virtual controller is globally stable, and concludes the first part of the proof.} 
    To analyse the finite-time stability, we solve the differential equation given by the virtual controller component-wise.
    \begin{align}
        &\frac{\mathrm{d}s_{1i}}{\mathrm{dt}} = -\frac{\Lambda}{t_{\mathrm{go}}}s_{1i} \nonumber\\
        \Rightarrow &\ln{s_{1i}} = \Lambda \ln{t_{\mathrm{go}}} + C_1 \nonumber\\
        % \Rightarrow &s_{1i} = s_{1i0}\left({{t_{\mathrm{go}}}/{t_f}}\right)^{\Lambda};\:\Dot{s}_{1i} = -({\Lambda s_{1i0}}/{t_f})\left({{t_{\mathrm{go}}}/{t_f}}\right)^{\Lambda - 1}\label{eq: s1,s1dot}
        \Rightarrow&\begin{array}{ll}
            s_{1i} = s_{1i0}\left({\frac{t_{\mathrm{go}}}{t_f}}\right)^{\Lambda}\\
            \Dot{s_{1i}} = -\frac{\Lambda}{t_f}s_{1i0}\left({\frac{t_{\mathrm{go}}}{t_f}}\right)^{\Lambda - 1}
        \end{array} \label{eq: s1,s1dot}
    \end{align}
    {where $C_1 = \ln{{s_{1i0}}/{t_f^\Lambda}}$.}
    The sliding surface and its derivative can be made finite-time convergent at $t = t_f$ by setting $\Lambda > 1$, and thus concluding the proof of this theorem.
\end{proof}

\subsection{MSS Design: Surface 2}\label{subsec:MSS2}
% At the beginning of powered descent stage, due to a variety of reasons, such as disturbances caused due to model uncertainty and atmospheric perturbations, the relation $\Dot{\mathbf{s}}_1 = -\frac{\Lambda}{t_{\mathrm{go}}}\mathbf{s}_1$ is not satisfied.
{At the beginning of powered descent stage, the relation $\Dot{\mathbf{s}}_1 = -\frac{\Lambda}{t_{\mathrm{go}}}\mathbf{s}_1$ is, in general, not satisfied due to initial conditions. Moreover, even if this condition is nearly satisfied, then also atmospheric perturbations and model uncertainties may drive the states away from this condition.}
The guidance law should be designed to drive $\Dot{\mathbf{s}}_1$ from its initial condition to $\Dot{\mathbf{s}}_1 = -\frac{\Lambda}{t_{\mathrm{go}}}\mathbf{s}_1$, maintain it there regardless of any disturbances, and eventually drive $\mathbf{s}_1$ to zero. To achieve this, a second sliding surface is proposed as:
\begin{align}
    \mathbf{s}_2 = \Dot{\mathbf{s}}_1 + \frac{\Lambda}{t_{\mathrm{go}}}\mathbf{s}_1 \label{eq: surf2}
\end{align}
From $\mathbf{s}_1$ we have $\Dot{\mathbf{s}}_1 = \mathbf{v} - {\mathbf{v}_f^d}$ and $\Ddot{\mathbf{s}}_1 = \mathbf{a}_c + \mathbf{g} - \Dot{\mathbf{v}}_f^d + \mathbf{a}_\mathrm{p}$, which when substituted in time derivative of \eqref{eq: surf2} gives:
\begin{align}
    \Dot{\mathbf{s}}_2 &= \Ddot{\mathbf{s}}_1 +\frac{\Lambda}{t_{\mathrm{go}}}\Dot{\mathbf{s}}_1 + \frac{\Lambda}{t_{\mathrm{go}}^2}\mathbf{s}_1 \nonumber\\    
     \Dot{\mathbf{s}}_2 &= \mathbf{a}_c + \mathbf{g} - \Dot{\mathbf{v}}_f^d + \frac{\Lambda}{t_{\mathrm{go}}}( \mathbf{v} - {\mathbf{v}_f^d} ) + \frac{\Lambda}{t_{\mathrm{go}}^2}( \mathbf{r} - {\mathbf{r}_f^d} + \mathbf{a}_\mathrm{p}) \label{eq: s2dot}
\end{align}
Since $\Dot{\mathbf{s}}_2$, as shown in \eqref{eq: s2dot}, has the acceleration term, the relative degree of $\mathbf{s}_2$ is 1, an appropriately chosen guidance law can drive $\mathbf{s}_2$ to zero. 

\begin{theorem}\label{thm:proposedGuidanceLaw}
    \textnormal{Consider the system dynamics in \eqref{eq: dynamics} and the sliding surface $\mathbf{s}_2$ given by \eqref{eq: surf2}, the guidance law: 
    \begin{align} \label{eq: accMain}
         \mathbf{a}_c = \mathbf{a}_{\mathrm{OTALG}} - \bm{\Phi}\sign{\mathbf{s}_2} - \mathbf{g}
    \end{align}
    where $\mathbf{a}_\mathrm{OTALG}$ is given by \eqref{eq:aOptimal}, will drive the second sliding surface $\mathbf{s}_2$ to zero in finite time, where $\bm{\Phi} = \mathrm{diag}\left\{\Phi_x,\ \Phi_y,\ \Phi_z \right\}$ is the sliding parameter, which depends on $d_i$ and $t_{\mathrm{go}}$.}
\end{theorem}
\begin{proof}
    We begin the proof by choosing the candidate Lyapunov function as $V_2 = 0.5\mathbf{s}_2^\mathrm{T}\mathbf{s}_2$. From direct observation, $V_2$ is positive definite everywhere except at origin where it is zero and is radially unbounded everywhere in the domain of $\mathbf{s}_2$ that is $\mathbb{R}^3$. To prove global stability we analyse the negative definiteness of the time derivative of $V_2$, given by $\Dot{V}_2 = \mathbf{s}_2^\mathrm{T}\Dot{\mathbf{s}}_2$. From \eqref{eq: s2dot} and \eqref{eq: accMain}, we get:
    \begin{align}
        \Dot{V}_2 &= \mathbf{s}_2^\mathrm{T}\Dot{\mathbf{s}}_2\nonumber\\
        \Dot{V}_2 &= \mathbf{s}_2^\mathrm{T} \left(\mathbf{a}_{\mathrm{OTALG}} - \bm{\Phi}\sign{\mathbf{s}_2} - \Dot{\mathbf{v}}_f^d + \frac{\Lambda}{t_{\mathrm{go}}}( \mathbf{v} - {\mathbf{v}_f^d} ) + \frac{\Lambda}{t_{\mathrm{go}}^2}( \mathbf{r} - {\mathbf{r}_f^d} ) + \mathbf{a}_\mathrm{p} \right). \label{eq: V2dotAcc}
    \end{align}
    From the nature of the landing guidance problem considered here, we have ${\mathbf{r}_f^d} = \mathbf{0}$ and ${\mathbf{v}_f^d} = \mathbf{0}$, and consequently $\Dot{\mathbf{v}}_f^d = \mathbf{0}$. With these considerations, from \eqref{eq: surf2}, we have:
    \begin{align}
        \mathbf{s}_2 = \mathbf{v} + \frac{\Lambda}{t_{\mathrm{go}}}\mathbf{r}. \label{eq: surf2New}
    \end{align}
    Then, from \eqref{eq:zeroeffort}, \eqref{eq:aOptimal} and \eqref{eq: V2dotAcc}, we get:
    {\begin{align}    
        % \Dot{V}_2 = \mathbf{s}_2^\mathrm{T}\bigg[ \frac{\Lambda - 6}{t_{\mathrm{go}}^2}\mathbf{r} &+ \frac{\Lambda - 4}{t_{\mathrm{go}}}\mathbf{v} + \mathbf{p}\frac{t_\mathrm{go}^2}{12} - \bm{\Phi}\sign\mathbf{s}_2 \bigg]. \label{eq: V2dotSubbed}
        \Dot{V}_2 = \mathbf{s}_2^\mathrm{T}\left[ \left(\frac{\Lambda - 6}{t_{\mathrm{go}}^2} \mathbf{r} + \frac{\Lambda - 4}{t_{\mathrm{go}}} \mathbf{v}\right) + \mathbf{a}_\mathrm{p} + \mathbf{p}\frac{t_{\mathrm{go}}^2}{12} - \bm{\Phi}\sign\mathbf{s}_2 \right]. \label{eq: V2dotSubbed}
    \end{align}}
    For $\Lambda = 2,\, 3$, from \eqref{eq: surf2New} and \eqref{eq: V2dotSubbed}, we have:
    \begin{align}
        \Dot{V}_2 = \frac{\Lambda - 4}{t_{\mathrm{go}}}\mathbf{s}_2^\mathrm{T}\mathbf{s}_2 + \mathbf{s}_2^\mathrm{T}\bigg( \mathbf{p}\frac{t_\mathrm{go}^2}{12} - \bm{\Phi}\sign\mathbf{s}_2 + \mathbf{a}_\mathrm{p}\bigg). \label{eq: V2dotFin}
    \end{align}
    We observe that the first term in \eqref{eq: V2dotFin}, for $\Lambda = 2,\, 3$, is negative. To ensure the second term is negative as well, $\bm{\Phi}$ must be chosen suitably. Examining the second term component-wise for negative semi-definiteness, that is $s_{2i}\cdot\left(p_i\frac{t_{\mathrm{go}}^2}{12} - \Phi_i \sign s_{2i} + a_{\mathrm{p}_i} \right) \leq 0$. This implies,
    \begin{align}
        % s_{2i}\left(p_i\frac{t_{\mathrm{go}}^2}{12} - \Phi_i \sign s_{2i} \right) &\leq 0 \nonumber\\
        s_{2i}\left( p_i\frac{t_{\mathrm{go}}^2}{12} + a_{\mathrm{p}_i} \right) &\leq \Phi_i \vert s_{2i}\vert \nonumber\\
        % \Rightarrow \Phi_i &\geq  \vert p_i \vert\frac{t_{\mathrm{go}}^2}{12}. \label{eq: phiNOap}
        \Phi_i \geq \left\vert p_i \frac{t_{\mathrm{go}}^2}{12} + a_{\mathrm{p}_i} \right\vert. \label{eq: Phi_i(initial)}
        % \label{eq: phiNOap}
        % . \label{eq: phiIneq}
    \end{align}
    where $a_{\mathrm{p}_i}$ is upper-bounded as $\vert a_{\mathrm{p}_i} \vert \leq a_{\mathrm{p}_{\mathrm{MAX}}}$.
    Thus, setting $\Phi_i$ according to the inequality \eqref{eq: Phi_i(initial)} will make \eqref{eq: V2dotFin} negative semi-definite and guarantee global asymptotic stability. Further, if the constraint on the sliding parameter in \eqref{eq: Phi_i(initial)} is satisfied with strict inequality, then $\Dot{V}_2$ will strictly be less than zero, and $V_2$ will converge to zero in finite time. Therefore, $\mathbf{s}_2$ also has finite time convergence, completing the proof for this theorem.
\end{proof}

% \subsection{Robustness Analysis}\label{subsec:RobustAnalysis}
% Now, we consider the disturbances in the environment, as expressed in \eqref{eq: dynamics}, that can cause the spacecraft to deviate from its nominal trajectory. Equation (\ref{eq: V2dotSubbed}) now becomes:
% \begin{align}
%     % \Dot{V}_2 = \mathbf{s}_2^\mathrm{T}\big[ {(\Lambda - 6)\mathbf{r}}/{t_{\mathrm{go}}^2} &+ {(\Lambda - 4)\mathbf{v}}/{t_{\mathrm{go}}}\nonumber\\ 
%     %     &+ \mathbf{a}_\mathrm{p} + \mathbf{p}{t_{\mathrm{go}}^2}/{12} - \bm{\Phi}\sign\mathbf{s}_2 \big]\\
%     \Dot{V}_2 = \mathbf{s}_2^\mathrm{T}\left[ \left(\frac{\Lambda - 6}{t_{\mathrm{go}^2}} \mathbf{r} + \frac{\Lambda - 4}{t_{\mathrm{go}^2}} \mathbf{v}\right) + \mathbf{a}_\mathrm{p}+ \mathbf{p}\frac{t_{\mathrm{go}^2}}{12} - \bm{\Phi}\sign\mathbf{s}_2 \right]
%     % \begin{array}{rl}
%     %      \Dot{V}_2 = & \mathbf{s}_2^\mathrm{T}\left[ \left(\frac{\Lambda - 6}{t_{\mathrm{go}^2}} \mathbf{r} + \frac{\Lambda - 4}{t_{\mathrm{go}^2}} \mathbf{v}\right) + \mathbf{a}_\mathrm{p} \right.\\
%     %      & \left.+ \mathbf{p}\frac{t_{\mathrm{go}^2}}{12} - \bm{\Phi}\sign\mathbf{s}_2 \right] 
%     % \end{array}.
% \end{align}
% Using a similar analysis used in the proof for Theorem \ref{thm:proposedGuidanceLaw}, we get the following condition for guaranteed robustness:
% \begin{align} \label{eq: Phi_i(initial)}
%     \Phi_i \geq \left\vert p_i \frac{t_{\mathrm{go}}^2}{12} + a_{\mathrm{p}_i} \right\vert
% \end{align}
% where $a_{\mathrm{p}_i}$ is upper-bounded as $\vert a_{\mathrm{p}_i} \vert \leq a_{\mathrm{p}_{\mathrm{MAX}}}$.

\section{ON THE CHOICE OF SLIDING PARAMETER}\label{sec:ChoiceOfPhi}
A common practice in sliding mode control is to fix the sliding parameter to be constant based on \textit{a-priori} obtained estimates of disturbances. However, the lower bound on $\Phi_i$ defined in \eqref{eq: Phi_i(initial)} depends on the states and the $t_\mathrm{go}$, and thus the sliding parameter should also be defined in the same manner. 
{The maximum value of RHS in \eqref{eq: Phi_i(initial)} can be determined and used as a sliding parameter, but it is an aggressive choice. 
An aggressive sliding parameter improves terminal precision as it commands a higher acceleration to turn towards the sliding surface as quickly as possible in the state-space and then maintain the states close to the sliding surfaces, requiring a higher control effort. Contrary to this, the states must come out of the vicinity of the sliding surface to avoid any collision with the terrain. To avoid aggressive sliding mode control and still achieve good accuracy in precision soft landing, we propose the sliding parameter as:
\begin{align} \label{eq:phi_i_FINAL}
    \Phi_i = k_1\left \vert p_i \right \vert \frac{t_{\mathrm{go}}^2}{12} + k_2a_{\mathrm{p}_{\mathrm{MAX}}}
\end{align}
where, $k_1,\ k_2$ are tunable positive constants.
Now, this $k_1,\, k_2$ can be tuned to achieve a reasonable trade-off between terrain avoidance and precision in soft landing.
Setting $k_1,\,k_2 = 1$ guarantees \eqref{eq: Phi_i(initial)} is satisfied. However, this requires divert manoeuvre to be executed with thrust higher than what is actually necessary. Setting $k_1,\,k_2 < 1$ may, in fact, be sufficient to successfully execute the divert manoeuvre while still maintaining the robustness in precision soft landing. This, however, may lead to violations of \eqref{eq: Phi_i(initial)}, leading to loss of finite time stability and guaranteed robustness.
We now show that even if the sliding constraint in \eqref{eq: Phi_i(initial)} is violated, that is by setting $k_1,\,k_2 < 1$, the guidance law in \eqref{eq: accMain} can still maintain robustness in precision soft landing using the concept of practical fixed-time stability (PFTS).}
PFTS, as stated in Lemma \ref{lem: PFTS}, is a relaxation of the fixed-time stability. The classical fixed-time stability requires the states to converge to the origin exactly. 
However with PFTS, this requirement is relaxed and the states are allowed to converge to an $\epsilon$-set containing the origin of the system. 
This relaxation expands the choice of sliding parameters and guarantees convergence in a fixed time as long as the Lyapunov function is provably decreasing.
To this end, we first define the duration for which the dominant divert manoeuvre is active, and prove that the duration of the dominant divert manoeuvre is finite. Finally, we prove that with the sliding parameter chosen in \eqref{eq:phi_i_FINAL}, practical fixed time stability can be achieved.

\begin{defn} [Duration of dominant divert manoeuvres] \label{def:divertMan}
    \textnormal{{Recalling from the OTALG law \eqref{eq:aOptimal} and noting that $\mathbf{p}{t_{\mathrm{go}}^2}/12$ term is responsible for the divert maneuver, and $({6}/{t_{\mathrm{go}}^2})\mathbf{ZEM} - ({2}/{t_{\mathrm{go}}})\mathbf{ZEV}$ term is responsible for the landing guidance.} Following this, the dominant divert manoeuvre is said to begin when $\vert p_i \vert t_{\mathrm{go}}^2/12 > \vert (6/t_{\mathrm{go}}^2)\mathrm{ZEM}_i - (2/t_{\mathrm{go}})\mathrm{ZEV}_i \vert$ and it ends when $\vert p_i \vert t_{\mathrm{go}}^2/12 < \vert (6/t_{\mathrm{go}}^2)\mathrm{ZEM}_i - (2/t_{\mathrm{go}})\mathrm{ZEV}_i \vert$. Further, in some cases when only a small divert acceleration is required, the dominant divert manoeuvre starts even with $\vert p_i \vert t_{\mathrm{go}}^2/12 < \vert (6/t_{\mathrm{go}}^2)\mathrm{ZEM}_i - (2/t_{\mathrm{go}})\mathrm{ZEV}_i \vert$ if $\sign{\Dot{d}_i}$ changes. In such cases, the dominant divert manoeuvre ends when $\sign{\Dot{d}_i}$ changes again.}
\end{defn}

\begin{figure}
  \centering
  \begin{tikzpicture}[scale=1]
    \begin{axis}[
      xlabel=$d_i(\mathrm{m})$,
      ylabel=$p_i(\mathrm{m}/\mathrm{s}^4)$,
      xmin=0,
      xmax=450,
      ymin=0,
      ymax=2.5,
      grid=both,
      grid style={line width=.1pt, draw=gray!10},
      major grid style={line width=.2pt,draw=gray!50},
      axis lines=middle,
      ]
      \addplot[domain=0:500, samples=500, red, line width=1.5pt] {9500*500*((x*exp(-9500/(1+x^2)))/(x^2 + 1)^2)};
    \end{axis}
  \end{tikzpicture}
  \caption{Behaviour of $p_i$ with respect to $d_i$ for $l_{1,i} = 1,\,l_{2,i}=9500,\,l_{3,i}=500$.}
  \label{fig: abs_pi}
\end{figure}
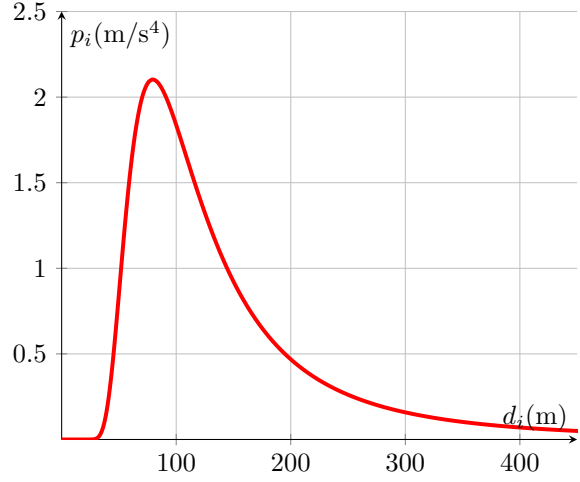

\begin{rem}\label{rem:analysisOfDivTerm}
    \textnormal{From the analysis of $p_i$ in Section 4.3 of the paper by \cite{basar2023fueloptimal}, the behaviour of $\vert p_i \vert$ with respect to $d_i$ as shown in Fig. \ref{fig: abs_pi} is a decreasing function for all $\vert d_i \vert > \vert d_i^* \vert$ where $d_i^*$ is the value of $d_i$ for which $p_i$ is maximum, that is, $\vert p_i(d_i^*) \vert = p_\mathrm{MAX}$, given as:
    \begin{align} \label{eq:d_crit}
        d_i^* = \pm\frac{\sqrt{\sqrt{l_{2,i}^2 - 2l_{1,i}l_{2,i} + 4l_{1,i}^2} + l_{2,i} - l_{1,i}}}{\sqrt{3}}.
    \end{align}
    Also, note from the divert term in \eqref{eq:aOptimal} and its definition in \eqref{eq: p_i}, for $d_i>0$ we have,
    \begin{align*} %\label{eq:pi_dot}
        % \frac{\partial (p_i\frac{t_\mathrm{go}^2}{12})}{\partial d_i} =& \frac{\Gamma t_\mathrm{go}^2}{12}[ 1 - {4d_i^2}/{(d_i^2 + l_{1,i})} + {2l_{2,i}d_i^2}/{(d_i^2 + l_{1,i})^2} ]
        \frac{\partial (p_it_\mathrm{go}^2/12)}{\partial d_i} =& \Gamma \left[ 1 - \frac{4d_i^2}{(d_i^2 + l_{1,i})} + \frac{2l_{2,i}d_i^2}{(d_i^2 + l_{1,i})^2} \right] \frac{t_\mathrm{go}^2}{12}.
    \end{align*}
    where $\Gamma = \frac{l_{2,i}l_{3,i}e^{-\psi_i}}{(d_i^2 + l_{1,i})^2} > 0$ and $\frac{t_\mathrm{go}^2}{12} \geq 0$. The above expression can be then made negative for all values of $d_i > d_i^*$ by suitably choosing $l_{1,i}$ and $l_{2,i}$. This implies that when the lander approaches the barriers, that is, $d_i$ decreases, with initial conditions $d_i > d_i^*$, the divert term will necessarily increase and therefore will cause the dominant divert manoeuvre to begin. Similar logic also holds true for $d_i < 0$.}
    % {Note that the case $\vert d_i \vert < \vert d_i^* \vert$ cannot happen as this would mean that the lander has crossed the barrier and crashed into the terrain. Hence this case is not discussed here.}

    \textnormal{However, the lander's initial position may be such that $\vert d_i \vert < \vert d_i^* \vert$ for a certain selection of tunable parameters in \eqref{eq:d_crit}. In such scenario, it may be noted that if the initial velocity vector is pointed away from the terrain, due to the motion of the lander, $d_i$ will momentarily increase which will then strengthen the divert thrust due to increasing $p_i$. In this situation, the lander will move away from the terrain and successfully avoid crashing. However, if the initial velocity direction is towards the terrain, $d_i$ decreases which causes the divert thrust to weaken due to decreasing $p_i$. In this situation, the lander will not be able to move away from the terrain, and may eventually crash. }

    \textnormal{An undesirable situation like this can be avoided by tuning the constants $l_{1,i},\ l_{2,i}$ in \eqref{eq:d_crit} to reduce the value of $d_i^*$ such that $\vert d_i \vert > \vert d_i^* \vert$. However, from \eqref{eq:aOptimal} and \eqref{eq: p_i} we can observe that changing $l_{1,i},\ l_{2,i}$ may adversely effect the guidance law, and hence deteriorate the overall performance. Thus, it is important that the terminal landing phase of soft landing, be initiated sufficiently far away from the terrains.
}
\end{rem}

\begin{proposition} \label{thm:DurationOfDivertMan}
    \textnormal{The duration of a dominant divert manoeuvre is finite.}
\end{proposition}
\begin{proof}    
Consider that the dominant divert manoeuvre begins for some $\vert d_i \vert > \vert d_i^* \vert$. During this time, the larger divert acceleration implies that the rate at which the lander approaches the barriers reduces. If the maximum thrust that the lander can generate is sufficiently large, then the sign of velocity, that is the direction of the velocity vector, will change and the lander will start to move away from the barrier. Observe that, in \eqref{eq:aOptimal} when $t_{\mathrm{go}}$ is large and $\vert d_i \vert \gg \vert d_i^* \vert$, the magnitude of ZEM/ZEV component is of the order of $O(1/t_{\mathrm{go}} + g_i)$, and when the $t_{\mathrm{go}}$ is small, the magnitude of ZEM/ZEV component is of the order of $O(1/t_{\mathrm{go}}^2)$. Finally, for the mid-ranges of $t_{\mathrm{go}}$, the magnitude of ZEM/ZEV component has the order of $O(g_i)$. Since initially $\vert d_i \vert \gg \vert d_i^* \vert$, $\vert p_i  \vert t_{\mathrm{go}}^2/12 \approx 0$. However, as $t$ increases, $t_{\mathrm{go}}$ decreases and $\lim_{t_{\mathrm{go}}\to 0}\vert p_i  \vert t_{\mathrm{go}}^2/12 = 0$. This implies that at the end of the landing mission the magnitude of divert acceleration term in \eqref{eq:aOptimal} comes sufficiently close to zero which is less than the magnitude of ZEM/ZEV term, and the dominant divert manoeuvre comes to an end in finite time. Since the mission is a fixed final time mission, this also implies that any dominant divert manoeuvre will come to an end in finite time.
\end{proof}

\begin{theorem}\label{thm:PFTS}
    \textnormal{The trajectory of the lander, governed by the dynamics given by \eqref{eq: dynamics} and under the guidance law $\mathbf{a}_c = \mathbf{a}_{\mathrm{OTALG}} - \bm{\Phi}\sign \mathbf{s}_2 - \mathbf{g}$ as defined in \eqref{eq: accMain}, have practical fixed-time stability (PFTS).}
\end{theorem}
\begin{proof}
From \eqref{eq: V2dotFin} we can obtain $\Dot{V}_2 = \sum \Dot{V}_{2i},\: i = x,\,y,\,z$ where:
\begin{align}
    % \Dot{V}_{2i} &= ({\Lambda - 4}) s_{2i}^2/{t_{\mathrm{go}}} + s_{2i}(p_i{t_\mathrm{go}^2}/{12} - \Phi_i\sign s_{2i}) \nonumber\\
    \Dot{V}_{2i} &= \left(\frac{\Lambda - 4}{t_{\mathrm{go}}}\right) s_{2i}^2 + s_{2i}\left( p_i\frac{t_\mathrm{go}^2}{12} - \Phi_i\sign s_{2i}\right) \nonumber\\
    \Rightarrow \Dot{V}_{2i} &= A + B %\left(\frac{\Lambda - 4}{t_{\mathrm{go}}}\right) \vert  s_{2i} \vert^2  - \Phi_i\vert  s_{2i} \vert + s_{2i}p_i\frac{t_\mathrm{go}^2}{12}. 
    \label{eq:V2Dots2Norm}
\end{align}
where, $\mathrm{A} = \left({\Lambda - 4}\right) \frac{\vert  s_{2i} \vert^2}{t_{\mathrm{go}}}  - \Phi_i\vert  s_{2i} \vert$ and $\mathrm{B} = s_{2i}p_i\frac{t_\mathrm{go}^2}{12}$. Then, for $\Lambda=2,\,3$,
% \begin{align}
%     \mathrm{A} &= \left(\frac{\Lambda - 4}{t_{\mathrm{go}}}\right) \left(2V_{2i}\right)  - \Phi_i\vert  s_{2i} \vert \nonumber\\
%     \Rightarrow \mathrm{A} &\leq \left(\frac{\Lambda - 4}{t_{\mathrm{go}}}\right) \left(2V_{2i}\right)  - \Phi_i\left(\sqrt{2}V_{2i}^{1/2}\right) \nonumber\\
%     \Rightarrow \mathrm{A} &\leq -\sqrt{2}\Phi_iV_{2i}^{1/2}. \label{eq: A}
% \end{align}
\begin{align}
    \mathrm{A} = 2\frac{\Lambda - 4}{t_{\mathrm{go}}}V_{2i} -\sqrt{2}\Phi_iV_{2i}^{1/2} \leq -\sqrt{2}\Phi_iV_{2i}^{1/2}. \label{eq: A}
\end{align}
Similarly, for $B$, from Lemma \ref{lem: YoungsIneq}:
\begin{align}
    \mathrm{B} &= s_{2i}p_i\frac{t_\mathrm{go}^2}{12} \nonumber\\
    % &\leq {((2V_{2i})^{\frac{1+p_1}{2}} + p_1\vert p_i \vert^{\frac{1+p_1}{p_1}})t_\mathrm{go}^2}/{12(1+p_1)}\\\label{eq: B}
    % \mathrm{B} &= s_{2i}p_i\frac{t_\mathrm{go}^2}{12} \leq \bigg(\frac{2^{\frac{1+p_1}{2}}}{1+p_1}V_{2i}^{\frac{1+p_1}{2}} + \frac{p_1}{1+p_1}\vert p_i \vert^{\frac{1+p_1}{p_1}}\bigg)\frac{t_\mathrm{go}^2}{12}\label{eq: B}
    % \Rightarrow\mathrm{B} &\leq \left(\frac{1}{1+p_1}\vert s_{2i} \vert^{1+p_1} + \frac{p_1}{1+p_1}\vert p_i \vert^{\frac{1+p_1}{p_1}}\right)\frac{t_\mathrm{go}^2}{12} \nonumber\\
    \Rightarrow \mathrm{B} &\leq \left(\frac{2^{\frac{1+p_1}{2}}}{1+p_1}V_{2i}^{\frac{1+p_1}{2}} + \frac{p_1}{1+p_1}\vert p_i \vert^{\frac{1+p_1}{p_1}}\right)\frac{t_\mathrm{go}^2}{12}. \label{eq: B}
\end{align}
where $p_1 > 0$.
% \begin{align}
%     \mathrm{B} &= s_{2i}p_i\frac{t_\mathrm{go}^2}{12} \nonumber\\
%     \Rightarrow\mathrm{B} &\leq \left(\frac{1}{1+p_1}\vert s_{2i} \vert^{1+p_1} + \frac{p_1}{1+p_1}\vert p_i \vert^{\frac{1+p_1}{p_1}}\right)\frac{t_\mathrm{go}^2}{12} \nonumber\\
%     \Rightarrow \mathrm{B} &\leq \left(\frac{2^{\frac{1+p_1}{2}}}{1+p_1}V_{2i}^{\frac{1+p_1}{2}} + \frac{p_1}{1+p_1}\vert p_i \vert^{\frac{1+p_1}{p_1}}\right)\frac{t_\mathrm{go}^2}{12}. \label{eq: B}
% \end{align}
Using \eqref{eq: A} and \eqref{eq: B} in \eqref{eq:V2Dots2Norm}, we get:
\begin{align}
    % &\begin{array}{lr}
    %      \Dot{V}_{2i} &\leq -\sqrt{2}\Phi_iV_{2i}^{1/2} + \left(\frac{2^{\frac{1+p_1}{2}}}{1+p_1}V_{2i}^{\frac{1+p_1}{2}} \right.\\
    %      &\left.+\frac{p_1}{1+p_1}\vert p_i \vert^{\frac{1+p_1}{p_1}}\right)\frac{t_\mathrm{go}^2}{12} 
    % \end{array}\\
    &\begin{array}{lr}\label{eq: comparetolem2}
         \Dot{V}_{2i} &\leq -\bigg(\sqrt{2}\Phi_iV_{2i}^{1/2} - \frac{2^{\frac{1+p_1}{2}}}{1+p_1}V_{2i}^{\frac{1+p_1}{2}}\frac{t_{\mathrm{go}}^2}{12}\bigg) + \frac{p_1}{1+p_1}\vert p_i \vert^{\frac{1+p_1}{p_1}}\frac{t_{\mathrm{go}}^2}{12}
    \end{array}
\end{align}
Using the results of Lemma \ref{lem: PFTS} in the Appendix (which gives the fundamental result on PFTS), we compare \eqref{eq: comparetolem2} with \eqref{eq: VdotPFS} to get
\begin{align}
    % \alpha_i &= \sqrt{2}\Phi_i;,\: \beta_i = -{2^{\frac{(1+p_1)}{2}}t_{\mathrm{go}}^2}/{12(1+p_1)} \\
    % \eta_i &= (p_1\vert p_i \vert^{\frac{1+p_1}{p_1}}t_{\mathrm{go}}^2)/{12(1+p_1)}\\
    % % \frac{p_1}{1+p_1}\vert p_i \vert^{\frac{1+p_1}{p_1}}\frac{t_{\mathrm{go}}^2}{12} \\
    % k &= 1,\: l = 1/2,\: m = {(1+p_1)}/{2}
    \begin{array}{rl}
        \alpha_i &= \sqrt{2}\Phi_i \\
        \beta_i &= -\frac{2^{\frac{(1+p_1)}{2}}}{1+p_1}\frac{t_{\mathrm{go}}^2}{12} \\
        \eta_i &= \frac{p_1}{1+p_1}\vert p_i \vert^{\frac{1+p_1}{p_1}}\frac{t_{\mathrm{go}}^2}{12}\\
        % \frac{p_1}{1+p_1}\vert p_i \vert^{\frac{1+p_1}{p_1}}\frac{t_{\mathrm{go}}^2}{12} \\
        k = 1,\: l = 1/2,\: &m = \frac{(1+p_1)}{2}
    \end{array}
\end{align}
where the conditions set by Lemma \ref{lem: PFTS} enforce that $p_1 > 1$. Then, the settling time is given as:
\begin{align}
    % T_i \leq \frac{\sqrt{2}}{\Phi_i\theta} - \frac{\sqrt{2}(1+p_1)}{2^{\frac{p_1}{2}}(p_1-1)\theta}\frac{12}{t_{\mathrm{go}}^2} \label{eq: settlingTimeComps}
    T_i \leq \frac{2}{\sqrt{2}\Phi_i\theta} - \frac{2(1+p_1)}{2^{\frac{p_1+1}{2}}(p_1-1)\theta}\frac{12}{t_{\mathrm{go}}^2} \label{eq: settlingTimeComps}
\end{align}
where $i\: =\: x,\,y,\,z$. The RHS of \eqref{eq: settlingTimeComps} must be strictly positive and less than $t_f$, which lead to the following conditions respectively:
{\begin{align} 
        \Phi_i &\leq \mathrm{L} \triangleq \frac{2^{\frac{p_1}{2}}(p_1-1)}{1+p_1}\frac{t_{\mathrm{go}}^2}{12}\label{eq: L}\\ 
        \Phi_i &> \mathrm{M} \triangleq \frac{1}{\frac{1+p_1}{2^{\frac{p_1}{2}}(p_1-1)}\frac{12}{t_{\mathrm{go}}^2} + \frac{t_f\theta}{\sqrt{2}}}.\label{eq: M}
\end{align}}
In the condition given by \eqref{eq: M}, we observe that when $t_{\mathrm{go}}$ is large, $\Phi_i \geq M \approx \sqrt{2}/(t_f\theta)$, which can be satisfied by choosing a sufficiently large $a_{p_{\mathrm{MAX}}}$ in \eqref{eq:phi_i_FINAL}. Further, when $t_{\mathrm{go}}$ is very small, $\Phi_i \geq M \approx 0$ is trivially satisfied.
Substituting the proposed sliding parameter \eqref{eq:phi_i_FINAL} in \eqref{eq: L} and rearranging the terms with $\vert p_i \vert = p_\mathrm{MAX}$, we get:
\begin{align}\label{eq:phi_iInL}
    k_1p_\mathrm{MAX} + \frac{12k_2a_{\mathrm{p}_{\mathrm{MAX}}}}{t_\mathrm{go}^2} \leq \frac{2^{\frac{p_1}{2}}(p_1-1)}{(1+p_1)}.
\end{align}
For all values of $t_\mathrm{go} \in (0,t_f]$, there exists a $p_1 > 1$ that satisfies \eqref{eq:phi_iInL}.
Further, in the condition given by \eqref{eq: M}, we observe that when $t_{\mathrm{go}}$ is large, $\Phi_i \geq M \approx \sqrt{2}/(t_f\theta)$, which can be satisfied by choosing a sufficiently large $a_{p_{\mathrm{MAX}}}$ in \eqref{eq:phi_i_FINAL}, and when $t_{\mathrm{go}}$ is very small, $\Phi_i \geq M \approx 0$ is trivially satisfied.
Therefore, the proposed sliding parameter satisfies the conditions for PFTS set by \eqref{eq: L} and \eqref{eq: M}, with settling time bounded by \eqref{eq: settlingTimeComps}.
Thus, even when the global finite time stability of Theorem \ref{thm:proposedGuidanceLaw} is not satisfied, the trajectories of the lander are PFTS, with settling time bounded by \eqref{eq: settlingTimeComps}.
\end{proof}

\section{SIMULATIONS AND DISCUSSIONS}\label{sec:Simulations}
\subsection{Simulation Setup and Parameters}\label{subsec:SimsSetupParams}
To demonstrate the effectiveness of MSS-OTALG, results from simulations are presented in this section.
We assume a point-mass lander with specific impulse $I_{\mathrm{sp}} = 225$ s, $T_{\mathrm{max}} = 31000$ N. The value of $T_{\mathrm{max}}$ has been chosen based on the necessary condition derived in Section 4.3 of the paper by \cite{basar2023fueloptimal}. The desired terminal states are ${\mathbf{r}_f^d} = [0,\, 0,\, 0]^\mathrm{T}$ m, ${\mathbf{v}_f^d} = [0,\, 0,\, 0]^\mathrm{T}$ m/s.
% , at terminal time $t_f = 100$ s. 
The terminal time $(t_f)$ plays a big role in obtaining fuel-optimal guidance law. However, calculating optimal $t_f$ is challenging. A method to determine the optimal $t_f$ was presented by \cite{Guo_CTVG_2012}, however this does not consider the time lost in the divert manoeuvres required for avoiding collision. Another method presented by \cite{Guo_Hawkins_Wie_2013} uses a line search to determine optimal $t_f$ which is extremely computationally expensive. 
In the paper by \cite{Zhang_Guo_Ma_Zeng_2017}, $t_f = 100$s was chosen for terrain avoidance in $z$-axis, which is greater than the feasible minimum $t_f$ required for soft landing with thrust-limited engines:
\begin{equation}
    t_{f_\mathrm{min}} = \max\left\{ -\frac{v_z(0)}{a_\mathrm{max} - g},\, \frac{-v_z(0) + \sqrt{v_z^2(0) - 2(a_\mathrm{max} - g)r_z(0)}}{a_\mathrm{max} - g} \right\},
\end{equation}
In the simulation studies presented in this paper, for terrain avoidance in all three axes, we follow the same $t_f = 100$s, which serves as a more restrictive choice of terminal time. 
The simulation is stopped when $r_z = 0.05$ m or desired terminal time is achieved.

To emulate a trench surrounding the landing site on Mars, we consider the terrain that can be modelled as a $2$-step, flat-top shape (similar to the illustration in Fig. \ref{fig:barrierPlot}). The height and width of each step from the origin are given as $h_{i,i} = 500,\,h_{i,2} = 1000,\, w_{i,1} = 600,\, w_{i,2} = 1000$m.
To design the barriers, we  choose $\theta_{3} = 0.05^{\circ}$, with $\lambda_{i,2} = 6$, and $\lambda_{i,1} = 20$. The guidance law constant are chosen as $l_{1,i} = 1,\ l_{2,i} = 9500,\ $ and $l_{3,i} = 500$, which gives the margin of safety for the vertical motion barrier as $\delta = 1.2 \cdot d_{i}^* = 95.5$m. The local gravity at Mars is assumed to be $\textbf{g} = [0,\ 0,\, -3.7114]^\mathrm{T}$ m/s\textsuperscript{2}, and acceleration due to gravity on Earth $g_e = 9.807$ m/s\textsuperscript{2} (\cite{Acikmese_Ploen_2007}). Thruster actuation latency is incorporated as first-order delays as $\Dot{\mathbf{a}}_c = (\mathbf{a}_{c_\mathrm{ideal}} - \mathbf{a}_c)/\tau$ (\cite{Wibben_Furfaro_2016}) where $\tau = 0.0556$s to emulate 90\% step response in 50ms (\cite{Dawson_Brewster_Conrad_Kilwine_Chenevert_Morgan_2007}). In reality, the thrust commanded in never the exact thrust generated, especially in the case of solid rocket motors. To emulate this, we perturb the thrust command by $\pm 5\%$ using MATLAB's \texttt{rand()} command. Finally, for sliding mode control we utilise $\Phi_i = k_1\vert p_i \vert {t_{\mathrm{go}}^2}/{12} + k_2a_{p_{\mathrm{MAX}}}$ where $k_1 = 0.8$ and $k_2 = 0.2$. To avoid the chattering problem associated with the signum function, we use the saturation function with boundary layer width $\epsilon = 0.1$.
\subsection{Illustration of a Numerical Example}\label{subsec: nominal}
\begin{figure*}
    \centering
    \def\svgwidth{0.94\linewidth}
    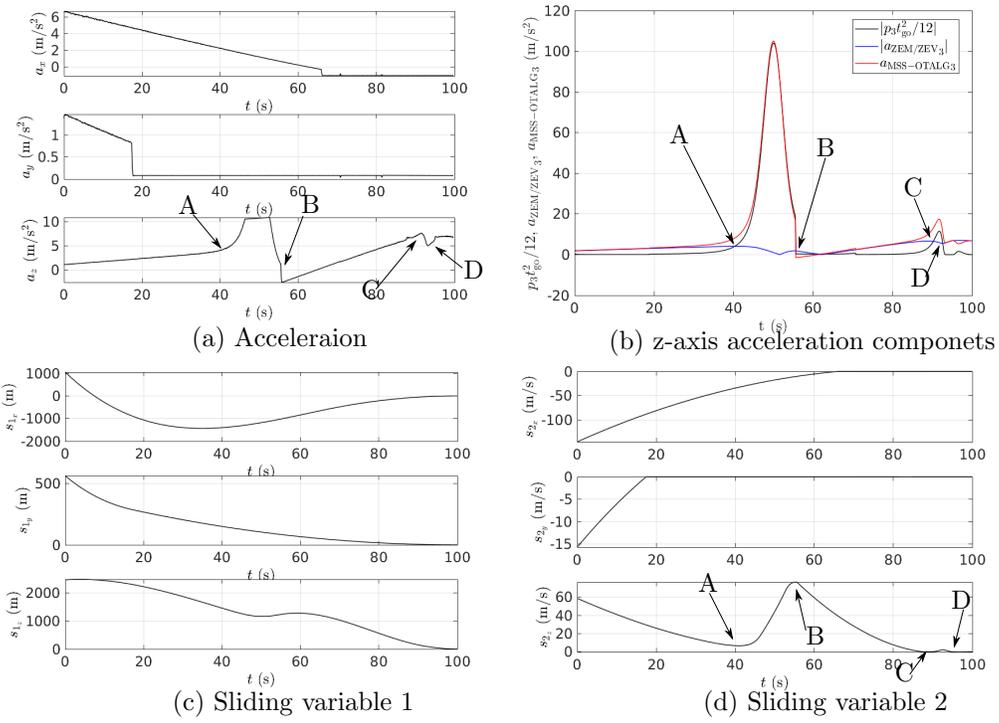
    \caption{Trajectory, velocity, commanded acceleration, divert term and sliding variables. [A: Divert manoeuvre 1 begins, B: Divert manoeuvre 1 ends, C: Divert manoeuvre 2 begins, D: Divert manoeuvre 2 ends]}
    \label{fig:bigPlot}
\end{figure*}
To showcase the nominal performance of the proposed guidance law, the simulation results are presented in Figure \ref{fig:bigPlot}. The initial conditions for this simulation are: ${\mathbf{r}_0} = [1051.86,\, 562.15,\, 2459.07]^\mathrm{T}$ m, ${\mathbf{v}_0} = [-165,\, -26.91,\, 9.45]^\mathrm{T}$ m/s and $m_0 = 1905$ kg. From the trajectories, position, velocity and the commanded acceleration plots in Figs. \ref{fig:bigPlot}a-d, it may be observed that the lander rises initially with a positive $v_z$ and then begins to descend at $t = 4$s under the influence of gravity. During this time, to slow down the descent, a positive $a_z$ is continued to be commanded by the guidance law. Besides, note that a large and negative $v_x$ causes the lander to overshoot the desired landing site along the $x$-direction, which prompts the guidance law to generate a large and positive $a_x$ in order to slow down the lateral motion and bring the lander towards the landing site.
As the lander nears the vertical motion barrier at $z = 1000$m (refer to Figs. \ref{fig:bigPlot}a-b), the first dominant divert manoeuvre begins at $t = 40.2$s as the divert term surpasses the ZEM/ZEV term in the overall acceleration command (refer to \eqref{eq:aOptimal}, \eqref{eq: accMain}), which is evident from Fig. \ref{fig:bigPlot}e. As the vertical motion barrier is encountered, the divert acceleration and hence the commanded acceleration ramps up smoothly and does not exhibit discontinuities in the $a_z$ profile, which is justified from \eqref{eq: p_i} and Fig. \ref{fig: abs_pi}. Meanwhile, under the influence of positive $a_x$, the lander crosses the $x = -1000$m mark at $t=55.5$s. At this point, the vertical motion barrier switches from $\rho_{z,3} = 1000 + \delta$ to $\rho_{z,2} = 500 + \delta$. As a consequence, the magnitude of divert term falls below the magnitude of ZEM/ZEV term, and the first dominant divert manoeuvre comes to an end (refer to Fig. \ref{fig:bigPlot}e). This also leads to the discontinuity observed in the $a_z$ profile at $t = 55.5$s. On the other hand, the small discontinuities observed in the $a_x$ and $a_y$ profiles (refer to Fig. \ref{fig:bigPlot}d) are due to $s_{2x}$ and $s_{2y}$ reaching nearly zero at around $t = 65.8$s and $t = 17.3$s, respectively, as can be seen in Fig. \ref{fig:bigPlot}g. Those time-instants onwards, very small magnitude of $a_x$ and $a_y$ are only commanded to maintain $s_{2x}$ and $s_{2y}$, respectively, close to zero. 

The effect of the divert term can also be observed in the sliding variables $s_{1z}$ and $s_{2z}$. Recall from \eqref{eq:phi_i_FINAL} that the rate at which the $s_{2z}$ converges to zero is dependent on the choice of $k_1$ and $k_2$ in the sliding parameter. Larger values of of $k_1$ and $k_2$ imply that the convergence of $s_{2z}$ to zero is more aggressive. However, to execute the divert manoeuvre whenever necessary, smaller value of $k_1$ and $k_2$ are desirable in order to allow the state-space trajectory to leave the neighbourhood of the sliding surface $s_{2z}=0$. Hence, $k_1 = 0.8,\, k_2 = 0.2$ are chosen in this simulation.

Similar to the first divert manoeuvre, as observed from Fig. \ref{fig:bigPlot}e, the second divert manoeuvre takes place when the lander approaches the last vertical motion barrier $\rho_{z,1} = \delta$ near the landing site. This slows down the lander further, thus further facilitating soft landing. However, the divert manoeuvre, in this case, ends soon due to small $t_\mathrm{go}$.
However, when the lander crosses this barrier, the sign of divert term changes, as expected from its expression in \eqref{eq: p_i}. To avoid this behaviour, it is recommended that $\rho_{z,1}$ be sufficiently close to $r_{fz}^d$.

In this numerical example, there are two dominant divert manoeuvres. Since the number of divert manoeuvres is finite, and the sliding variables reach zero in finite time (see Figs. \ref{fig:bigPlot}f-g), robustness of the MSS-OTALG is validated following the notion of PFTS in Theorem \ref{thm:PFTS}.

\subsection{Comparative Simulation Study}\label{subsec:TerrainAvoid}
Very few papers in existing literature have addressed both the problems of precision soft landing and terrain avoidance simultaneously. In this regard, it may be noted that it was stated by \cite{Wibben_Furfaro_2016}, a widely-referred precision soft landing paper, that the OSG presented therein could be augmented with the method by \cite{Zhou_Xia_2014} for achieving precision soft-landing with terrain avoidance. This augmentation methods was later improved by \cite{Zhang_Guo_Ma_Zeng_2017}. Also, the MSS-OTALG presented in this paper is an expansion over the OTALG by \cite{basar2023fueloptimal}, which also dealt with both these problems in an integrated way. Thus, in this section, the simulation study in Section \ref{subsec: nominal} is extended to incorporate a comparative analysis of the performance of the MSS-OTALG w.r.t. that of the OTALG presented by \cite{basar2023fueloptimal} and augmented OSG (\cite{Wibben_Furfaro_2016, Zhang_Guo_Ma_Zeng_2017}). Illustrative examples of this comparative study using the same initial conditions and desired terminal condition, as in the previous subsection, under zero and non-zero atmospheric perturbation are presented in Figs. \ref{fig:traj_vel} (in Section \ref{subsubsec:zero_aP}) and \ref{fig:traj_vel_aP} (in Section \ref{subsubsec:non-zero_aP}), respectively. Subsequently, extensive comparison study results under both zero and non-zero atmospheric perturbation are presented in Figs. \ref{fig:landingDispersion_0aP} and \ref{fig:landingDispersion_aP}, respectively. 

\subsubsection{Comparison study under zero atmospheric perturbations} \label{subsubsec:zero_aP}
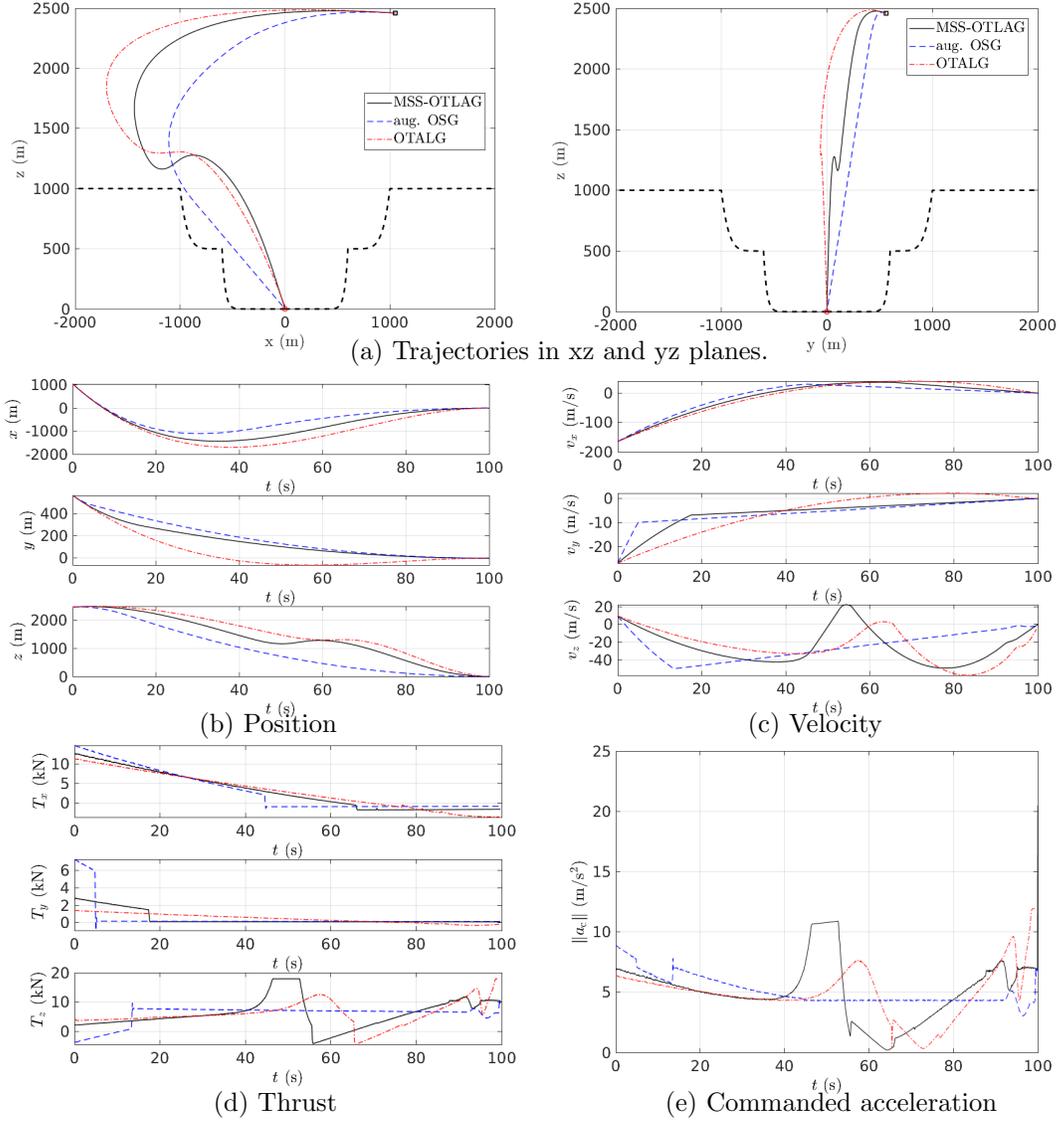
\begin{figure*}
    \centering
    \def\svgwidth{\linewidth}
    %% Creator: Inkscape 1.3.2 (091e20e, 2023-11-25, custom), www.inkscape.org
%% Accompanies image file 'trajCompPlots_final.pdf' (pdf, eps, ps)
\begingroup%
  \makeatletter%
  \providecommand\color[2][]{%
    \errmessage{(Inkscape) Color is used for the text in Inkscape, but the package 'color.sty' is not loaded}%
    \renewcommand\color[2][]{}%
  }%
  \providecommand\transparent[1]{%
    \errmessage{(Inkscape) Transparency is used (non-zero) for the text in Inkscape, but the package 'transparent.sty' is not loaded}%
    \renewcommand\transparent[1]{}%
  }%
  \providecommand\rotatebox[2]{#2}%
  \newcommand*\fsize{\dimexpr\f@size pt\relax}%
  \newcommand*\lineheight[1]{\fontsize{\fsize}{#1\fsize}\selectfont}%
  \ifx\svgwidth\undefined%
    \setlength{\unitlength}{554.99997116bp}%
    \ifx\svgscale\undefined%
      \relax%
    \else%
      \setlength{\unitlength}{\unitlength * \real{\svgscale}}%
    \fi%
  \else%
    \setlength{\unitlength}{\svgwidth}%
  \fi%
  \global\let\svgwidth\undefined%
  \global\let\svgscale\undefined%
  \makeatother%
  \begin{picture}(1,1.04438782)%
    \lineheight{1}%
    \setlength\tabcolsep{0pt}%
    \put(0,0){\includegraphics[width=\unitlength,page=1]{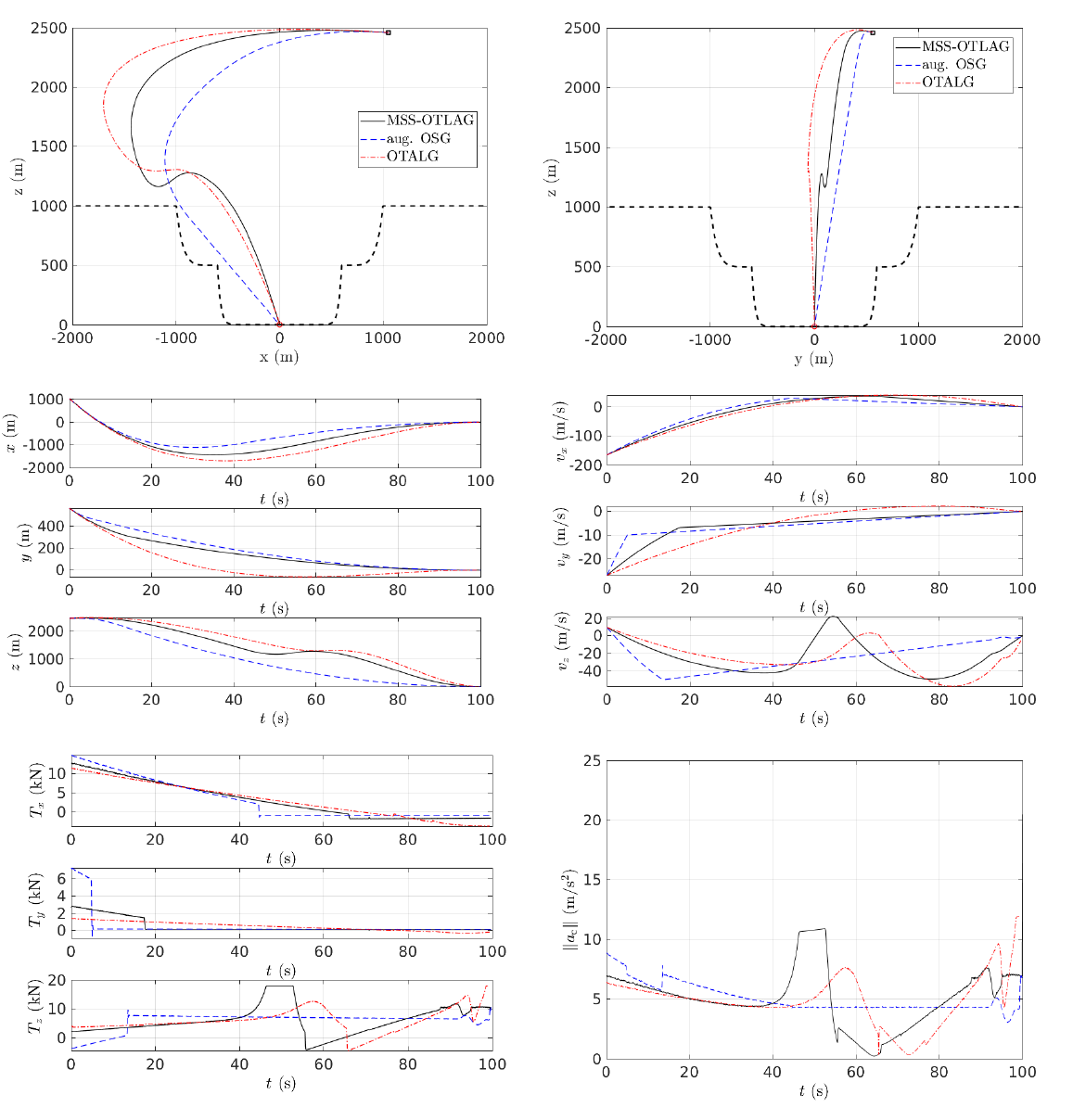}}%
    \put(0.76270036,0.00384689){\color[rgb]{0,0,0}\makebox(0,0)[t]{\lineheight{1.25}\smash{\begin{tabular}[t]{c}(e) Commanded acceleration\end{tabular}}}}%
    \put(0.25131638,0.00384689){\color[rgb]{0,0,0}\makebox(0,0)[t]{\lineheight{1.25}\smash{\begin{tabular}[t]{c}(d) Thrust\end{tabular}}}}%
    \put(0.74795214,0.35249567){\color[rgb]{0,0,0}\makebox(0,0)[t]{\lineheight{1.25}\smash{\begin{tabular}[t]{c}(c) Velocity\end{tabular}}}}%
    \put(0.24455962,0.35249567){\color[rgb]{0,0,0}\makebox(0,0)[t]{\lineheight{1.25}\smash{\begin{tabular}[t]{c}(b) Position\end{tabular}}}}%
    \put(0.51231772,0.69573914){\color[rgb]{0,0,0}\makebox(0,0)[t]{\lineheight{1.25}\smash{\begin{tabular}[t]{c}(a) Trajectories in xz and yz planes.\end{tabular}}}}%
  \end{picture}%
\endgroup%

    \caption{Comparison of MSS-OTALG, aug. OSG and OTALG under no atmospheric perturbations.}
    \label{fig:traj_vel}
\end{figure*}
\begin{figure*}
    \centering
    \def\svgwidth{\linewidth}
    %% Creator: Inkscape 1.1.2 (0a00cf5339, 2022-02-04), www.inkscape.org
%% Accompanies image file 'MC_final.pdf' (pdf, eps, ps)
\begingroup%
  \makeatletter%
  \providecommand\color[2][]{%
    \errmessage{(Inkscape) Color is used for the text in Inkscape, but the package 'color.sty' is not loaded}%
    \renewcommand\color[2][]{}%
  }%
  \providecommand\transparent[1]{%
    \errmessage{(Inkscape) Transparency is used (non-zero) for the text in Inkscape, but the package 'transparent.sty' is not loaded}%
    \renewcommand\transparent[1]{}%
  }%
  \providecommand\rotatebox[2]{#2}%
  \newcommand*\fsize{\dimexpr\f@size pt\relax}%
  \newcommand*\lineheight[1]{\fontsize{\fsize}{#1\fsize}\selectfont}%
  \ifx\svgwidth\undefined%
    \setlength{\unitlength}{1397.77521552bp}%
    \ifx\svgscale\undefined%
      \relax%
    \else%
      \setlength{\unitlength}{\unitlength * \real{\svgscale}}%
    \fi%
  \else%
    \setlength{\unitlength}{\svgwidth}%
  \fi%
  \global\let\svgwidth\undefined%
  \global\let\svgscale\undefined%
  \makeatother%
  \begin{picture}(1,0.22090887)%
    \lineheight{1}%
    \setlength\tabcolsep{0pt}%
    \put(0,0){\includegraphics[width=\unitlength,page=1]{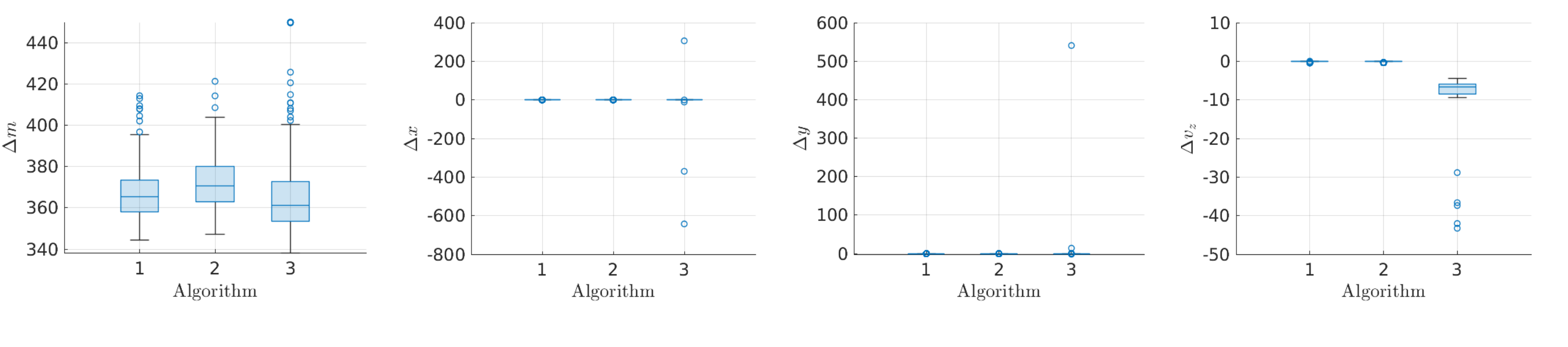}}%
    \put(0.50397518,0.00143321){\color[rgb]{0,0,0}\makebox(0,0)[t]{\lineheight{1.25}\smash{\begin{tabular}[t]{c}(b), (c) Landing Precision in $x-$ and $y-$ direction\end{tabular}}}}%
    \put(0.13334244,0.00149986){\color[rgb]{0,0,0}\makebox(0,0)[t]{\lineheight{1.25}\smash{\begin{tabular}[t]{c}(a) Fuel Consumption\end{tabular}}}}%
    \put(0.87770038,0.00321607){\color[rgb]{0,0,0}\makebox(0,0)[t]{\lineheight{1.25}\smash{\begin{tabular}[t]{c}(d) Terminal Descent Velocity\end{tabular}}}}%
  \end{picture}%
\endgroup%

    \caption{MC simulation results for fuel consumption, landing dispersion and residual terminal velocity ($a_{\mathrm{P}} = 0$). [1: MSS-OTALG, 2: augmented OSG, 3: OTALG]}
    \label{fig:landingDispersion_0aP}
\end{figure*}

From the trajectories in Fig. \ref{fig:traj_vel}a, position profiles in Fig. \ref{fig:traj_vel}b and velocity profiles in Fig. \ref{fig:traj_vel}c, it is observed that all three guidance laws under comparison are able to drive the lander towards the desired landing site precisely and softly, while also successfully avoiding the terrain. From the net acceleration profile in Fig. \ref{fig:traj_vel}e, observe that the augmented OSG applies a larger initial acceleration to bring the trajectory close to the sliding surface till $t = 44.6$s and subsequently a nearly constant acceleration to maintain the trajectory near the sliding surface.
However, as the augmented OSG has been formulated to avoid terrain only in the $z-$direction, it avoids the terrain only marginally in the $x-y$ direction, as can be observed in Fig. \ref{fig:traj_vel}a. On the other hand, the OTALG is formulated to avoid the terrain in any direction. When the lander is away from any terrain, OTALG behaves similar to the OGL \cite{Ebrahimi_Bahrami_Roshanian_2008} in the sense that it applies just enough acceleration for precision soft-landing. But, when the terrain is encountered, the divert term starts to dominate the ZEM/ZEV term in the guidance law, leading to a higher acceleration commanded by OTALG to avoid the terrain and again bring it back to the desired landing site when the terrain is sufficiently avoided. 
The consequence of these two very different acceleration profiles is that while the augmented OSG has an excellent performance in terms of landing precision but the OTALG outperforms the former in terms of fuel consumption and terrain avoidance. Further, in the case of augmented OSG, increasing the sliding parameter to improve precision also increases the turn rate of the lander, which may be detrimental to the sensitive equipment carried onboard. To this end, the MSS-OTALG developed in this paper finds a middle ground between these conflicting objectives. 

The presence of the sliding term in the MSS-OTALG imparts a high degree of precision, akin to the augmented OSG to the tune of $\Delta x_f = 9.31\cdot10^{-7}$m, $\Delta y_f = 3.64\cdot10^{-7}$m and $\Delta v_{zf} = -0.02$m/s for MSS-OTALG, 
$\Delta x_f = -9.98\cdot10^{-6}$m, $\Delta y_f = 3.45\cdot10^{-6}$m and $\Delta v_{zf} = -0.01$m/s for augmented OSG and 
$\Delta x_f = -7.32\cdot10^{-2}$m, $\Delta y_f = 3.0\cdot10^{-3}$m and $\Delta v_{zf} = -3.67$m/s for OTALG. Moreover, as the OTALG is also embedded in the formulation of MSS-OTALG, it inherits the feature of terrain avoidance in all directions and near-fuel-optimality from OTALG, which is evident from fuel consumption data ($\Delta m = 391.37$ kg for MSS-OTALG, $\Delta m = 394.60$ kg for aug. OSG and $\Delta m = 379.22$ kg for OTALG). In this way, MSS-OTALG reconciles seemingly conflicting objectives, with both the sliding and divert terms collaborating to achieve terrain-avoided soft landing objectives.
\begin{table}
\centering
\caption{Normal Distribution for Initial Conditions}\label{tab:ic}
\begin{tabular}{l p{.5cm} p{.5cm} p{.5cm} p{.5cm} p{.5cm} p{.5cm} p{.5cm}}
\hline
State & $x_0$ & $y_0$ & $z_0$ & $v_{\mathrm{x0}}$ & $v_{\mathrm{y0}}$ & $v_{\mathrm{z0}}$ & $m_0$ \\
\hline
Mean & 0 & 0 & 2500 & 0 & 0 & -80 & 1905 \\
SD & 2200 & 2200 & 400 & 80 & 80 & 20 & 0 \\
\hline
\end{tabular}
\end{table}

{Monte Carlo simulations are also conducted using 300 initial conditions selected from the normal distribution outlined in Table \ref{tab:ic}. The purpose is to objectively assess the soft-landing accuracy and fuel consumption statistics of the three guidance laws under comparison. The results of these simulations are depicted using box plot representation in Fig. \ref{fig:landingDispersion_0aP}, while the corresponding statistical data is summarised in Table \ref{tab:missValues}. Augmented OSG and MSS-OTALG exhibit a high degree of accuracy in precision soft landing (refer to Fig. \ref{fig:landingDispersion_0aP}b-d), yet the former consumes significantly more fuel than the latter (as determined via paired t-test with null hypothesis $H_0: \Delta m_{\mathrm{aug.\, OSG}} = \Delta m_\mathrm{{MSS-OTLAG}}$ against $H_1: \Delta m_{\mathrm{aug.\, OSG}} > \Delta m_\mathrm{{MSS-OTLAG}}$, which gives $t_{\mathrm{stat}} = 14.35$). This can also be observed from Fig. \ref{fig:landingDispersion_0aP}a. Conversely, OTALG is found to consume quite less fuel compared to MSS-OTALG, but performs poorly in terms of precision soft-landing performance. Thus, these Monte Carlo studies validate that the MSS-OTALG developed in this paper achieves significantly superior performance in precision soft-landing while also avoiding terrain yet demanding near-to-optimal fuel consumption.} 

\noindent\begin{minipage}{\linewidth}
\centering
\captionof{table}{Terminal States Statistics ($a_\mathrm{p} = 0$)}\label{tab:missValues}\resizebox{\linewidth}{!}{%
\begin{tabular}{c|c|c|c|c|c|c|c|c}
\hline
\multirow{2}{*}{Guidance Law} & \multicolumn{4}{c|}{Mean}                               & \multicolumn{4}{c}{SD}                                 \\ 
\cline{2-9}
                              & $\Delta m$   & $\Delta x$  & $\Delta y$  & $\Delta v_z$ & $\Delta m$ & $\Delta x$  & $\Delta y$  & $\Delta v_z$  \\ 
\hline
MSS-OTALG                     & $366.67$ & $1.65\cdot10^{-5}$ & $4.37\cdot10^{-5}$ & $-3.32\cdot10^{-2}$  & $12.72$  & $6.25\cdot10^{-4}$ & $6.48\cdot10^{-4}$ & $8.29\cdot10^{-2}$     \\
aug. OSG                      & $371.73$    & $1.67\cdot10^{-6}$   & $-1.55\cdot10^{-5}$ & $-6.69\cdot10^{-2}$   & $12.89$  & $6.78\cdot10^{-4}$  & $6.19\cdot10^{-4}$  & $0.13$     \\
OTALG                         & $365.91$   & $-2.40$  & $1.84$ & $-7.43$    & $19.62$  & $46.29$    & $31.23$    & $4.22$     
\end{tabular}
}
\end{minipage}

\subsubsection{Comparison study under non-zero atmospheric perturbations} \label{subsubsec:non-zero_aP}
\begin{figure*}
    \centering
    \def\svgwidth{\linewidth}
    %% Creator: Inkscape 1.3.2 (091e20e, 2023-11-25, custom), www.inkscape.org
%% Accompanies image file 'trajCompPlots_aP_final.pdf' (pdf, eps, ps)
\begingroup%
  \makeatletter%
  \providecommand\color[2][]{%
    \errmessage{(Inkscape) Color is used for the text in Inkscape, but the package 'color.sty' is not loaded}%
    \renewcommand\color[2][]{}%
  }%
  \providecommand\transparent[1]{%
    \errmessage{(Inkscape) Transparency is used (non-zero) for the text in Inkscape, but the package 'transparent.sty' is not loaded}%
    \renewcommand\transparent[1]{}%
  }%
  \providecommand\rotatebox[2]{#2}%
  \newcommand*\fsize{\dimexpr\f@size pt\relax}%
  \newcommand*\lineheight[1]{\fontsize{\fsize}{#1\fsize}\selectfont}%
  \ifx\svgwidth\undefined%
    \setlength{\unitlength}{552.75002307bp}%
    \ifx\svgscale\undefined%
      \relax%
    \else%
      \setlength{\unitlength}{\unitlength * \real{\svgscale}}%
    \fi%
  \else%
    \setlength{\unitlength}{\svgwidth}%
  \fi%
  \global\let\svgwidth\undefined%
  \global\let\svgscale\undefined%
  \makeatother%
  \begin{picture}(1,1.04510515)%
    \lineheight{1}%
    \setlength\tabcolsep{0pt}%
    \put(0,0){\includegraphics[width=\unitlength,page=1]{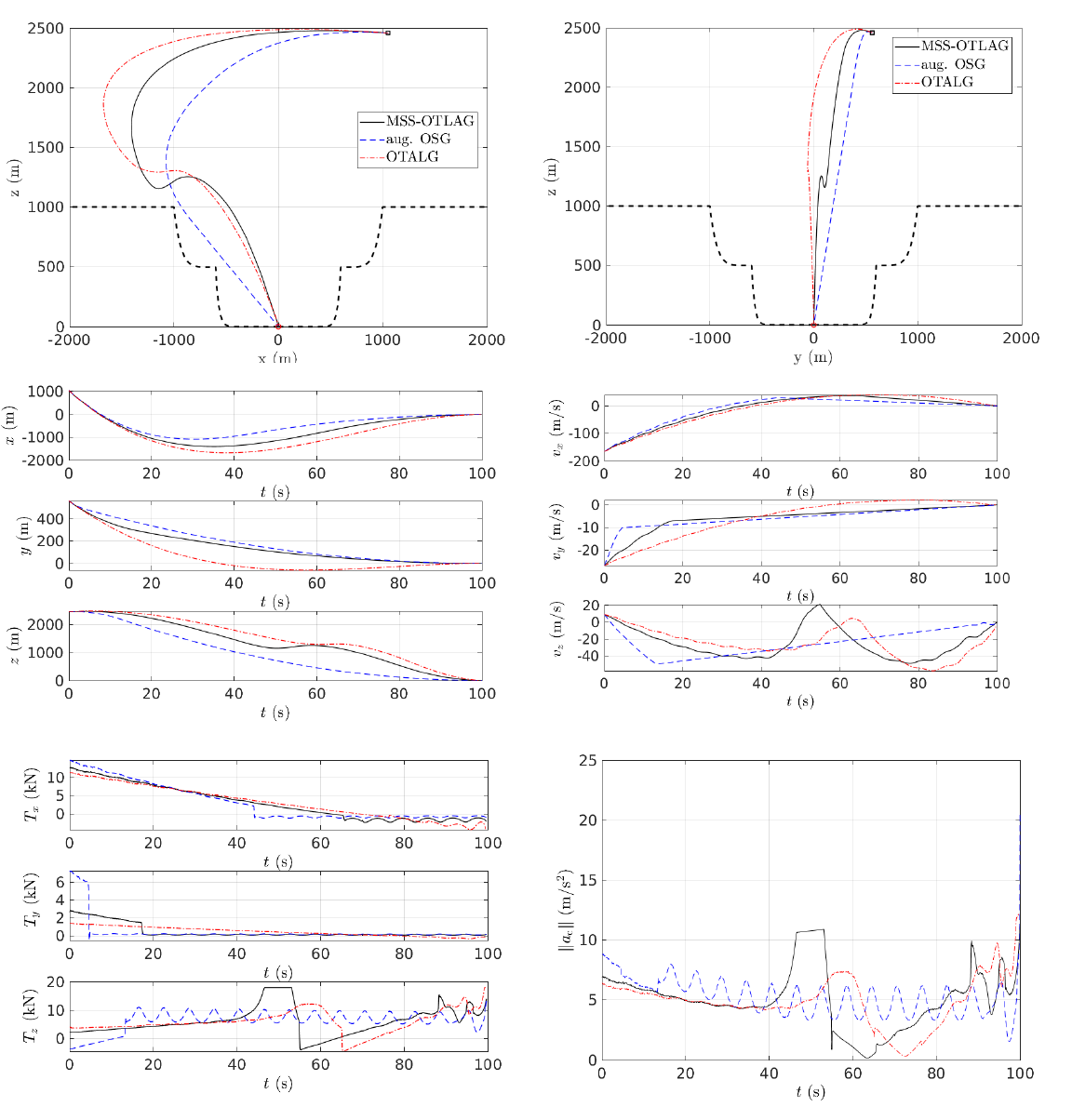}}%
    \put(0.76229094,0.00386254){\color[rgb]{0,0,0}\makebox(0,0)[t]{\lineheight{1.25}\smash{\begin{tabular}[t]{c}(e) Commanded acceleration\end{tabular}}}}%
    \put(0.24882539,0.00386254){\color[rgb]{0,0,0}\makebox(0,0)[t]{\lineheight{1.25}\smash{\begin{tabular}[t]{c}(d) Thrust\end{tabular}}}}%
    \put(0.74748272,0.36207151){\color[rgb]{0,0,0}\makebox(0,0)[t]{\lineheight{1.25}\smash{\begin{tabular}[t]{c}(c) Velocity\end{tabular}}}}%
    \put(0.24204113,0.35935781){\color[rgb]{0,0,0}\makebox(0,0)[t]{\lineheight{1.25}\smash{\begin{tabular}[t]{c}(b) Position\end{tabular}}}}%
    \put(0.51088912,0.70128473){\color[rgb]{0,0,0}\makebox(0,0)[t]{\lineheight{1.25}\smash{\begin{tabular}[t]{c}(a) Trajectories in xz and yz planes.\end{tabular}}}}%
  \end{picture}%
\endgroup%

    \caption{Comparison of MSS-OTALG, aug. OSG and OTALG under bounded atmospheric perturbations.}
    \label{fig:traj_vel_aP}
\end{figure*}
\begin{figure*}
    \centering
    \def\svgwidth{\linewidth}
    %% Creator: Inkscape 1.1.2 (0a00cf5339, 2022-02-04), www.inkscape.org
%% Accompanies image file 'MC_aP_final.pdf' (pdf, eps, ps)
\begingroup%
  \makeatletter%
  \providecommand\color[2][]{%
    \errmessage{(Inkscape) Color is used for the text in Inkscape, but the package 'color.sty' is not loaded}%
    \renewcommand\color[2][]{}%
  }%
  \providecommand\transparent[1]{%
    \errmessage{(Inkscape) Transparency is used (non-zero) for the text in Inkscape, but the package 'transparent.sty' is not loaded}%
    \renewcommand\transparent[1]{}%
  }%
  \providecommand\rotatebox[2]{#2}%
  \newcommand*\fsize{\dimexpr\f@size pt\relax}%
  \newcommand*\lineheight[1]{\fontsize{\fsize}{#1\fsize}\selectfont}%
  \ifx\svgwidth\undefined%
    \setlength{\unitlength}{1393.50002018bp}%
    \ifx\svgscale\undefined%
      \relax%
    \else%
      \setlength{\unitlength}{\unitlength * \real{\svgscale}}%
    \fi%
  \else%
    \setlength{\unitlength}{\svgwidth}%
  \fi%
  \global\let\svgwidth\undefined%
  \global\let\svgscale\undefined%
  \makeatother%
  \begin{picture}(1,0.21189878)%
    \lineheight{1}%
    \setlength\tabcolsep{0pt}%
    \put(0,0){\includegraphics[width=\unitlength,page=1]{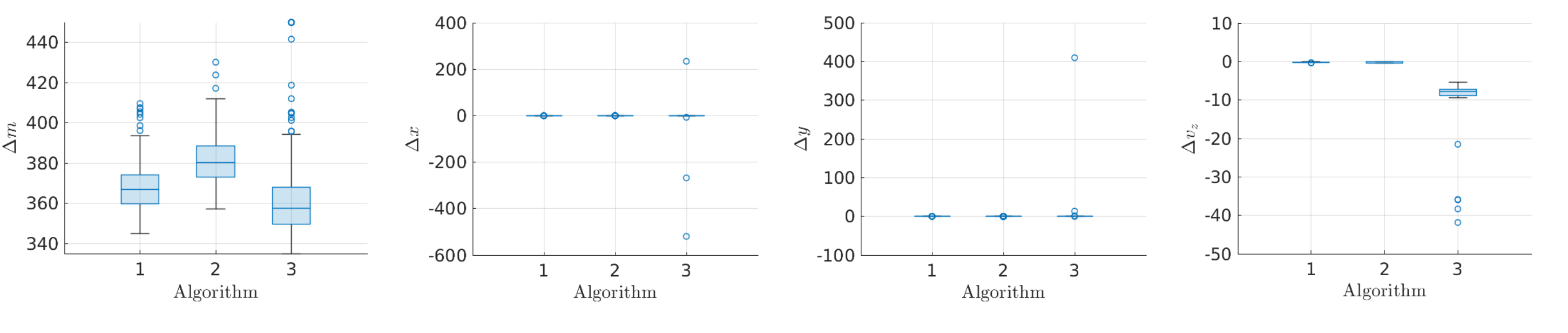}}%
    \put(0.49045138,0.00143761){\color[rgb]{0,0,0}\makebox(0,0)[t]{\lineheight{1.25}\smash{\begin{tabular}[t]{c}(b), (c) Landing Precision in $x-$ and $y-$ direction\end{tabular}}}}%
    \put(0.11868155,0.00150446){\color[rgb]{0,0,0}\makebox(0,0)[t]{\lineheight{1.25}\smash{\begin{tabular}[t]{c}(a) Fuel Consumption\end{tabular}}}}%
    \put(0.86532314,0.00322593){\color[rgb]{0,0,0}\makebox(0,0)[t]{\lineheight{1.25}\smash{\begin{tabular}[t]{c}(d) Terminal Descent Velocity\end{tabular}}}}%
  \end{picture}%
\endgroup%

    \caption{MC simulation results for fuel consumption, landing dispersion and residual terminal velocity ($a_{\mathrm{P}} \neq 0$). [1: MSS-OTALG, 2: augmented OSG, 3: OTALG]}
    \label{fig:landingDispersion_aP}
\end{figure*}
Presence of disturbances, such as those caused due to atmosphere, can cause the lander to go off the nominal course and can cause the lander to perform poorly in terms of precision soft landing. In this section, comparative simulation study under non-zero atmospheric perturbations is presented in which $\mathbf{a}_p(t) = 0.3\mathbf{a}_c\sin{\left( {\pi t}/{3} \right)}$ (\cite{Zhang_Guo_Ma_Zeng_2017}) is considered as the model for atmospheric perturbations. Using the same initial conditions as in Fig. \ref{fig:traj_vel}, illustrative examples using the three guidance laws - MSS-OTALG, augmented OSG and OTALG - are presented in Fig. \ref{fig:traj_vel_aP}. All of them are able to avoid the terrain and precisely and softly land close to the desired landing site, as evident from Fig. \ref{fig:traj_vel_aP}a-c. Similar to the zero perturbation case, the augmented OSG demands a higher initial acceleration due to the sliding term, which in effect improves the disturbance rejection capability as can be observed in Fig. \ref{fig:traj_vel_aP}d-e, where the command acceleration and thrust vary to attenuate the disturbances caused by the winds. This results in almost no oscillations in the velocity of the lander, as can be observed in Fig. \ref{fig:traj_vel_aP}c. On the other hand, since the OTALG does not have any disturbance rejection capability, the thrust and commanded acceleration profiles are similar to that of zero perturbation case, however this causes the lander to sway continuously and thus degrade the soft-landing accuracy, as can be observed in the velocity profiles shown in Fig. \ref{fig:traj_vel_aP}c. Coming to the MSS-OTALG, sufficient disturbance rejection can be observed due to the sliding term, and at the same time the terrain is avoided due to the divert term. Since the constants $k_1,\, k_2 < 1$ have been chosen for the sliding parameter, the disturbance rejection by MSS-OTALG is not as effective as that by the augmented OSG. However, it attenuates the disturbances caused by the wind significantly better that the OTALG. This phenomenon can be observed in the acceleration command and thrust profiles (refer to Fig. \ref{fig:traj_vel_aP}d-e). It can also be observed that as the divert term's influence increases, it dominates the effect of the sliding term to avoid the terrain. Then, when the divert term is small the sliding term operates to mitigate disturbances, improving the accuracy of the precision soft landing.

{To assess both the soft-landing precision and fuel usage in the presence of atmospheric disturbances, Monte-Carlo simulations has been conducted utilising the same 300 initial conditions as used in the Monte-Carlo simulations shown in Fig. \ref{fig:landingDispersion_0aP}. The outcomes are graphically represented via box plots in Fig. \ref{fig:landingDispersion_aP} and numerically summarised in Table \ref{tab:missValues_aP}. It's evident from the simulations that the augmented OSG, as anticipated from the thrust and acceleration profiles in Fig. \ref{fig:traj_vel_aP}d-e, consumes more fuel. However, it consistently delivers the most accurate precision soft-landings. MSS-OTALG exhibits slightly higher fuel consumption compared to OTALG but, showcases commendable performance in precision soft-landing akin to the augmented OSG. On the other hand, although OTALG demonstrates lower fuel consumption compared to the augmented OSG and MSS-OTALG, it suffers from significantly poorer performance in precision soft-landing. Hence, the Monte-Carlo simulations effectively validate the robustness of MSS-OTALG in mitigating the impact of non-zero atmospheric perturbations while achieving the main objectives of terrain-avoided precision soft landing.}

\noindent\begin{minipage}{\linewidth}
\centering
\captionof{table}{Terminal States Statistics ($a_\mathrm{p} \neq 0$)}\label{tab:missValues_aP}\resizebox{\linewidth}{!}{%
\begin{tabular}{c|c|c|c|c|c|c|c|c} 
\hline
\multirow{2}{*}{Guidance Law} & \multicolumn{4}{c|}{Mean}                            & \multicolumn{4}{c}{SD}                               \\ 
\cline{2-9}
                              & $\Delta m$ & $\Delta x$  & $\Delta y$ & $\Delta v_z$ & $\Delta m$ & $\Delta x$ & $\Delta y$ & $\Delta v_z$  \\ 
\hline
MSS-OTALG                     & $367.73$ & $2.75\cdot10^{-5}$ & $-4.71\cdot10^{-5}$ & $-0.17$  & $11.74$  & $1.37\cdot10^{-3}$ & $1.37\cdot10^{-4}$ & $4.49\cdot10^{-2}$     \\
aug. OSG                      & $381.33$    & $3.03\cdot10^{-5}$   & $-9.02\cdot10^{-5}$ & $-0.19$   & $12.46$  & $1.05\cdot10^{-3}$  & $1.02\cdot10^{-3}$  & $0.15$     \\
OTALG                         & $361.50$   & $-1.87$  & $1.41$ & $-8.21$    & $19.58$  & $36.36$    & $23.66$    & $3.71$     
\end{tabular}
}
\end{minipage}

\section{CONCLUSION}
To allow spacecrafts to safely land with high precision and low fuel consumption while avoiding terrain, a guidance law, named MSS-OTALG, using the recently developed Optimal Terrain Avoidance Landing Guidance Law (OTALG) and the concept of Multiple Sliding Surfaces (MSS) is presented in this paper.
% Expanding upon the Optimal Terrain Avoidance Landing Guidance Law (OTALG) recently presented in literature with the help of sliding mode control by multiple sliding surfaces (MSS), a near-fuel optimal and robust guidance law for precision soft-landing in hazardous terrain, named MSS-OTALG, is presented in this paper. 
The proposed guidance law inherits the near-fuel optimality and terrain avoidance features of the OTALG, and the incorporation of MSS renders the guidance law robust against disturbances as well.
To allow the lander to manoeuvre away from the terrain, a state and time-dependent sliding parameter is introduced, and practical fixed time stability is proven under the proposed guidance law. 
Finally, extensive computer simulations validate the ability of the MSS-OTALG to avoid the terrain and precisely and softly land at the desired landing site while having low fuel consumption, under realistic limitations posed by thruster dynamics, thrust constraints and atmospheric disturbances.
When compared against the OTALG and the augmented version of the optimal sliding guidance (OSG) using Monte Carlo simulations, it was observed that while MSS-OTALG consumes more fuel than the near-fuel optimal OTALG, it is able to consistently give better precision soft landing performance which is comparable to the augmented OSG, but at much lesser cost. Thus, the proposed guidance law succeeds in effectively finding the middle ground in terms of all the performance measures.
The results presented in this paper assume the local terrain near the desired landing site is known perfectly \textit{a-priori} and use that information to generate the terrain barriers beforehand.
Thus, to improve autonomy, it is necessary to integrate a terrain feature detection algorithm, that can ingest information from onboard sensors and use this to generate the terrain barriers in an online manner.
% However, this is never the case for planetary landing.
% As a point of future study, to improve the autonomy of the proposed guidance law, it is necessary to integrate a terrain feature detection algorithm, that can ingest information from onboard sensors and use this to generate the terrain barriers in an online manner.

\section{Appendix}
\begin{lemma} (Young's Inequality) \label{lem: YoungsIneq}
    For any vector $\mathbf{x}$, $\mathbf{y} \in \mathbb{R}^n$, $\mathbf{x}^{\mathrm{T}}\mathbf{y} \leq \Vert\mathbf{x}\Vert^a/a + \Vert\mathbf{y}\Vert^b/b$ holds true
    % \begin{equation}
    %     \mathbf{x}^{\mathrm{T}}\mathbf{y} \leq \Vert\mathbf{x}\Vert^a/a + \Vert\mathbf{y}\Vert^b/b \label{eq:young}
    % \end{equation}
    where $a,\,b>1$ and $(a-1)(b-1)=1$.
\end{lemma}

\begin{lemma}\label{lem: PFTS}\cite{Polyakov2012, Jiang_Hu_Friswell_2016}
    Consider the system, 
    \begin{equation}
        \Dot{\mathbf{x}} = \mathbf{f}(\mathbf{x}(t)),\: \mathbf{x}(t_0) = \mathbf{x}_0. \label{eq: nonlinearSys}
    \end{equation}
    Suppose there exists a Lyapunov function $V(\mathbf{x})$ such that,
    \begin{equation}
        \Dot{V}(\mathbf{x}) \leq -\left(\alpha V^l(\mathbf{x}) + \beta V^m(\mathbf{x})\right)^k + \eta \label{eq: VdotPFS}
    \end{equation}
    where $\alpha$, $\beta$, $l$, $m$, $k$ $> 0$, $lk < 1$, $mk > 1$ and $0<\eta<\infty$. Then the trajectories of \eqref{eq: nonlinearSys} are practically fixed-time stable, with residual set
    \begin{align}
            \left\{ \lim_{t\rightarrow T} \mathbf{x} \left\vert\right. V(\mathbf{x}) \leq \min\left\{ \begin{array}{cc}
            &\alpha^{-\frac{1}{l}}\left(\frac{\eta}{1-\theta^k}\right)^{\frac{1}{lk}},\\
            &\beta^{-\frac{1}{l}}\left(\frac{\eta}{1-\theta^k}\right)^{\frac{1}{mk}}
            \end{array}\right\}\right\}
         \label{eq: residualSet}
    \end{align}
    where $\theta \in (0,1]$ and the settling time, $T$ is upper bounded by
    \begin{equation}
        T \leq \left( \frac{1}{\alpha^k\theta^k(1-lk)} + \frac{1}{\beta^k\theta^k(mk-1)} \right). \label{eq: settlingtime}
    \end{equation}
\end{lemma}

Throughout the paper the signum function, $\sign (\cdot)$ is defined as:
\begin{equation}\label{eq:signum}
    \sign \Delta \triangleq \left\{\begin{array}{cc}
        -1 & \mathrm{if}\: \Delta < 0\\
        0 & \mathrm{if}\: \Delta = 0\\
        1 & \mathrm{if}\: \Delta > 0
    \end{array}\right..
\end{equation}
Further, consider $\mathbf{x}\in\mathbb{R}^n$, then 
\begin{equation}\label{eq:vecsignum}
    \sign \mathbf{x} \triangleq \left[\sign x_1,\,\sign x_2,\dots,\sign x_n\right]^{\mathrm{T}}.
\end{equation}

{\footnotesize\bibliography{references}}

\begin{thebibliography}{10}
\providecommand{\url}[1]{#1}
\csname url@samestyle\endcsname
\providecommand{\newblock}{\relax}
\providecommand{\bibinfo}[2]{#2}
\providecommand{\BIBentrySTDinterwordspacing}{\spaceskip=0pt\relax}
\providecommand{\BIBentryALTinterwordstretchfactor}{4}
\providecommand{\BIBentryALTinterwordspacing}{\spaceskip=\fontdimen2\font plus
\BIBentryALTinterwordstretchfactor\fontdimen3\font minus \fontdimen4\font\relax}
\providecommand{\BIBforeignlanguage}[2]{{%
\expandafter\ifx\csname l@#1\endcsname\relax
\typeout{** WARNING: IEEEtran.bst: No hyphenation pattern has been}%
\typeout{** loaded for the language `#1'. Using the pattern for}%
\typeout{** the default language instead.}%
\else
\language=\csname l@#1\endcsname
\fi
#2}}
\providecommand{\BIBdecl}{\relax}
\BIBdecl

\bibitem{Mars_2020}
\BIBentryALTinterwordspacing
O.~Yakimenko, ``Landing safely on earth, then the moon and mars,'' \emph{Aerospace America}, vol.~58, no.~11, pp. 30--30, dec 2020. [Online]. Available: \url{https://aerospaceamerica.aiaa.org/year-in-review/landing-safely-on-earth-then-the-moon-and-mars/}
\BIBentrySTDinterwordspacing

\bibitem{Acikmese_Ploen_2007}
B.~Acikmese and S.~R. Ploen, ``Convex programming approach to powered descent guidance for mars landing.'' \emph{Journal of Guidance, Control, and Dynamics}, vol.~30, p. 1353–1366, 2007.

\bibitem{Simplício_Marcos_Joffre_Zamaro_Silva_2018}
P.~Simplício, A.~Marcos, E.~Joffre, M.~Zamaro, and N.~Silva, ``Review of guidance techniques for landing on small bodies,'' \emph{Progress in Aerospace Sciences}, vol. 103, p. 69–83, nov 2018.

\bibitem{Chai_Tsourdos_Savvaris_Chai_Xia_Philip_Chen_2021}
R.~Chai, A.~Tsourdos, A.~Savvaris, S.~Chai, Y.~Xia, and C.~Philip~Chen, ``Review of advanced guidance and control algorithms for space/aerospace vehicles,'' \emph{Progress in Aerospace Sciences}, vol. 122, p. 100696, apr 2021.

\bibitem{Mao_Szmuk_Açıkmeşe_2016}
Y.~Mao, M.~Szmuk, and B.~Açıkmeşe, ``Successive convexification of non-convex optimal control problems and its convergence properties,'' in \emph{2016 IEEE 55th Conference on Decision and Control (CDC)}, 2016, p. 3636–3641.

\bibitem{Swaminathan_U.P_Ghose_2020}
S.~Swaminathan, R.~U.P, and D.~Ghose, ``Real time powered descent guidance algorithm for mars pinpoint landing with inequality constraints,'' in \emph{AIAA Scitech 2020 Forum}.\hskip 1em plus 0.5em minus 0.4em\relax American Institute of Aeronautics and Astronautics, jan 2020.

\bibitem{Byung_Jang_Hyung_1998}
B.~S. Kim, J.~G. Lee, and H.~S. Han, ``Biased {PNG} law for impact with angular constraint.'' \emph{IEEE Transactions on Aerospace and Electronic Systems}, vol.~34, p. 277–288, 1998.

\bibitem{Ebrahimi_Bahrami_Roshanian_2008}
B.~Ebrahimi, M.~Bahrami, and J.~Roshanian, ``Optimal sliding-mode guidance with terminal velocity constraint for fixed-interval propulsive maneuvers.'' \emph{Acta Astronautica}, vol.~62, p. 556–562, 2008.

\bibitem{Lu_2020}
P.~Lu, ``Theory of fractional-polynomial powered descent guidance,'' \emph{Journal of Guidance, Control, and Dynamics}, vol.~43, no.~3, p. 398–409, mar 2020.

\bibitem{Sánchez-Sánchez_Izzo_2018}
C.~Sánchez-Sánchez and D.~Izzo, ``Real-time optimal control via deep neural networks: Study on landing problems,'' \emph{Journal of Guidance, Control, and Dynamics}, vol.~41, no.~5, p. 1122–1135, may 2018.

\bibitem{You_Wan_Dai_Rea_2021}
S.~You, C.~Wan, R.~Dai, and J.~R. Rea, ``Learning-based onboard guidance for fuel-optimal powered descent,'' \emph{Journal of Guidance, Control, and Dynamics}, vol.~44, no.~3, p. 601–613, mar 2021.

\bibitem{Furfaro_Gaudet_Wibben_Kidd_Simo_2013}
R.~Furfaro, B.~Gaudet, D.~R. Wibben, J.~Kidd, and J.~Simo, ``Development of non-linear guidance algorithms for asteroids close-proximity operations.'' in \emph{AIAA Guidance, Navigation, and Control (GNC) Conference}, 2013.

\bibitem{Furfaro_Cersosimo_Wibben_2013}
R.~Furfaro, D.~Cersosimo, and D.~R. Wibben, ``Asteroid precision landing via multiple sliding surfaces guidance techniques.'' \emph{Journal of Guidance, Control, and Dynamics}, vol.~36, p. 1075–1092, 2013.

\bibitem{Gong_Guo_Lyu_Ma_Guo_2022}
Y.~Gong, Y.~Guo, Y.~Lyu, G.~Ma, and M.~Guo, ``Multi-constrained feedback guidance for mars pinpoint soft landing using time-varying sliding mode,'' \emph{Advances in Space Research}, vol.~70, no.~8, p. 2240–2253, oct 2022.

\bibitem{Wibben_Furfaro_2016}
D.~R. Wibben and R.~Furfaro, ``Optimal sliding guidance algorithm for mars powered descent phase.'' \emph{Advances in Space Research}, vol.~57, p. 948–961, 2016.

\bibitem{Bai_Guo_Zheng_2020}
C.~Bai, J.~Guo, and H.~Zheng, ``Optimal guidance for planetary landing in hazardous terrains,'' \emph{IEEE Transactions on Aerospace and Electronic Systems}, vol.~56, no.~4, p. 2896–2909, 2020.

\bibitem{Wang_Guo_Ma_Wie_2021}
P.~Wang, Y.~Guo, G.~Ma, and B.~Wie, ``Two-phase zero-effort-miss/zero-effort-velocity guidance for mars landing,'' \emph{Journal of Guidance, Control, and Dynamics}, vol.~44, no.~1, p. 75–87, jan 2021.

\bibitem{Gong_Guo_Ma_Zhang_Guo_2021}
Y.~Gong, Y.~Guo, G.~Ma, Y.~Zhang, and M.~Guo, ``Barrier lyapunov function-based planetary landing guidance for hazardous terrains.'' \emph{IEEE/ASME Transactions on Mechatronics}, vol.~27, p. 2764–2774, 2021.

\bibitem{Gong_Guo_Ma_Zhang_Guo_2022}
------, ``Prescribed performance-based powered descent guidance for step-shaped hazardous terrains.'' \emph{IEEE Transactions on Aerospace and Electronic Systems}, vol.~58, p. 1083–1095, 2022.

\bibitem{basar2023fueloptimal}
S.~Z. Basar and S.~Ghosh, ``Fuel-optimal powered descent guidance for hazardous terrain,'' \emph{IFAC-PapersOnLine}, vol.~56, no.~2, pp. 6018--6023, 2023, 22nd IFAC World Congress.

\bibitem{Polyakov2012}
A.~Polyakov, ``Nonlinear feedback design for fixed-time stabilization of linear control systems,'' \emph{{IEEE} Transactions on Automatic Control}, vol.~57, no.~8, pp. 2106--2110, Aug. 2012.

\bibitem{Jiang_Hu_Friswell_2016}
B.~Jiang, Q.~Hu, and M.~I. Friswell, ``Fixed-time attitude control for rigid spacecraft with actuator saturation and faults,'' \emph{IEEE Transactions on Control Systems Technology}, vol.~24, no.~5, p. 1892–1898, sep 2016.

\bibitem{Seeber_2003}
G.~Seeber, \emph{Satellite Geodesy}.\hskip 1em plus 0.5em minus 0.4em\relax Walter de Gruyter, 2003.

\bibitem{Zhou_Xia_2014}
L.~Zhou and Y.~Xia, ``Improved {ZEM/ZEV} feedback guidance for mars powered descent phase.'' \emph{Advances in Space Research}, vol.~54, p. 2446–2455, 2014.

\bibitem{Zhang_Guo_Ma_Zeng_2017}
Y.~Zhang, Y.~Guo, G.~Ma, and T.~Zeng, ``Collision avoidance {ZEM/ZEV} optimal feedback guidance for powered descent phase of landing on mars.'' \emph{Advances in Space Research}, vol.~59, p. 1514–1525, 2017.

\bibitem{Guo_CTVG_2012}
\BIBentryALTinterwordspacing
Y.~Guo, M.~Hawkins, and B.~Wie, ``Optimal feedback guidance algorithms for planetary landing and asteroid intercept.'' in \emph{AAS/AIAA astrodynamics specialist conference}, 2011, pp. 2011--588. [Online]. Available: \url{https://www.abe.iastate.edu/wp-content/blogs.dir/25/files/2011/09/AAS-11-588.pdf}
\BIBentrySTDinterwordspacing

\bibitem{Guo_Hawkins_Wie_2013}
------, ``Waypoint-optimized {Zero-Effort-Miss/Zero-Effort-Velocity} feedback guidance for mars landing.'' \emph{Journal of Guidance, Control, and Dynamics}, vol.~36, p. 799–809, 2013.

\bibitem{Dawson_Brewster_Conrad_Kilwine_Chenevert_Morgan_2007}
M.~Dawson, G.~Brewster, C.~Conrad, M.~Kilwine, B.~Chenevert, and O.~Morgan, ``Monopropellant hydrazine 700 lbf throttling terminal descent engine for mars science laboratory,'' in \emph{43rd AIAA/ASME/SAE/ASEE Joint Propulsion Conference and Exhibit}.\hskip 1em plus 0.5em minus 0.4em\relax American Institute of Aeronautics and Astronautics, jul 2007.

\end{thebibliography}

\end{document}